%% file: researchReportV6.tex
\documentclass[runningheads]{llncs}
\usepackage{lineno}
	\usepackage{fancyvrb}
	\usepackage{color}

	\usepackage{graphicx}
	\usepackage{amssymb,amsmath}
	\usepackage{enumerate}
	\usepackage[noend]{algpseudocode}
	\usepackage{algorithm}
	 \usepackage[colorlinks]{hyperref}
 
	\usepackage{multicol, tabularx, makecell, multirow, diagbox}
	
	\usepackage{marginnote}
        \usepackage{xspace}
		\usepackage{mathtools}

	\algdef{SE}[DOWHILE]{Do}{doWhile}{\algorithmicdo}[1]{\algorithmicwhile\ #1}

	\newcommand {\C}   {\mathbb C}
	\newcommand {\Z}   {\mathbb Z}
	\newcommand {\N}   {\mathbb N}

	\newcommand{\CYc}[1]{{\color{blue}Chee said: #1~}}
	
	\newcommand{\cored}[1]{{\color{red}#1}}
	\newcommand{\coblue}[1]{{\color{blue}#1}}
	\newcommand{\ass}{\leftarrow}
	
	\newcommand{\highlightModifs}[1]{{\color{blue}#1}}
	\renewcommand{\highlightModifs}[1]{{#1}}

	\newcommand{\Julialang}{\texttt{Julia}\xspace}
	\newcommand{\homfps}{\texttt{HOM4PS-2.0}\xspace}
	\newcommand{\bertini}{\texttt{Bertini}\xspace}
	\newcommand{\bertiniAMP}{\texttt{Bertini AMP}\xspace}
	\newcommand{\ccluster}{\texttt{ccluster}\xspace}
	\newcommand{\tcluster}{\texttt{tcluster}\xspace}
	
	\newcommand{\maple}{\texttt{Maple}\xspace}
	\newcommand{\HCfnJL}{\texttt{HomotopyContinuation.jl}\xspace}
	\newcommand{\HCJL}{\texttt{HomCont.jl}\xspace}
	\newcommand{\CCJL}{\texttt{Ccluster.jl}\xspace}
	
	\newcommand{\vecunkdim}[2]{#1_1,\ldots,#1_{#2}}
	
    \newcommand{\mvec}[1]{{\bf #1}}
    \newcommand{\mapf}{\mvec{f}}
    \newcommand{\mapg}{\mvec{g}}
    \newcommand{\maph}{\mvec{h}}
    \newcommand{\vecz}{\mvec{z}}
    \newcommand{\vecD}{\mvec{\Delta}}
    \newcommand{\vecB}{\mvec{B}}

    \newcommand{\veca}{\mvec{a}}
    \newcommand{\vecb}{\mvec{b}}
    \newcommand{\vecm}{\mvec{m}}
    \newcommand{\vecL}{\mvec{L}}
    \newcommand{\veczero}{\mvec{0}}
	\newcommand{\solsIn}[2]{\texttt{Zero}(#1,#2)}
    \newcommand{\nbSolsIn}[2]{\#(#1,#2)}
    \newcommand{\multipli}[2]{\#(#1,#2)}

    \newcommand{\TsTL}{$T_*^L$}

    \newcommand{\app}[1]{\widetilde{#1}}

    \newcommand{\mdim}{n}
    \renewcommand{\dim}{\mdim}
    \newcommand{\mdeg}{d}
    \renewcommand{\deg}{\mdeg}
    \newcommand{\last}[1]{\ensuremath{\dot{\mvec{#1}}}} 
    \newcommand{\others}[1]{\ensuremath{\overline{\mvec{#1}}}}
    \newcommand{\spec}[2]{(#1)_{#2}} 
    \newcommand{\contDisc}[1]{\Delta(#1)}
    
    \newcommand{\radius}[1]{r(#1)}
    
    \newcommand{\width}[1]{w(#1)}

	\newenvironment{prog}{\begin{tabbing}
	 xxxx\=xxxx\=xxxx\=xxxx\=xxxx\=xxxx\=xxxx\=xxxx\=xxxx\=xxxx\=xxxx\=xxxx\=xxxx\=
	\kill\\}{
		\end{tabbing}}
	\newcounter{indentcounter1}
	\newcounter{indentcounter2}
	\newcounter{indentcounter3}
	\newcommand{\Indent}[1][5]{
		\setcounter{indentcounter1}{#1}		
		\setcounter{indentcounter2}{1}		
		\setcounter{indentcounter3}{%
		    \value{indentcounter1}*\value{indentcounter2}}
		\hspace*{\value{indentcounter3} mm}} 	
	\newcommand{\lline}[1][0]{
	       \ \\ \Indent[#1]\texttt {} }
	
	\newenvironment{progb}[2][4]{ 
		\begin{center}
		  \fcolorbox{red}{white}{\begin{minipage}{0.75\textwidth}
	        	\vspace*{-#1\abovedisplayskip}
			\begin{prog}#2\end{prog}
	          \end{minipage}}
	        \end{center}
		}{}
	
	\newcommand{\bang}[1]{\langle #1\rangle}
	\newcommand{\moveToApp}[1]{}
	\newcommand{\intbox}{%
	  {\,\,\setlength{\unitlength}{.33mm}\framebox(4,7){}\,}}

	\newtheorem{propo}[lemma]{\textsc{Proposition}} %
	\newcommand{\bprol}[1]{\begin{propo} \label{pro:#1}}
	\newcommand{\eprol}{\end{propo}}
	\newcommand{\bproT}[2]{\begin{propo}[#1] \label{pro:#2}}
	\newcommand{\eproT}{\end{propo}}
	\newcommand{\refPro}[1]{Proposition~\protect{\ref{pro:#1}}}
	\newtheorem{coro}[lemma]{\textsc{Corollary}}
	\newcommand{\bcorT}[2]{\begin{coro}[#1] \label{cor:#2}}
	\newcommand{\ecorT}{\end{coro}}
	\newcommand{\beqarrys}{$$\begin{array}{llll}}
	\newcommand{\eeqarrys}{\end{array}$$}
	\newcommand{\btableTwo}[2][0.65]{\begin{center}
		\begin{tabular*}{#1\textwidth}{l @{\extracolsep{\fill}} lllll}
		#2
		\end{tabular*} \end{center}}
	\newcommand{\degree}{\mathop{deg}}	
	\newcommand{\ii}{\boldsymbol{i}}	
	\newcommand{\0}{\boldsymbol{0}}		
	\newcommand{\half}{\textstyle{\frac{1}{2}}} 
	\newcommand{\ib}{\subseteq }		
	\newcommand{\ignore}[1]{}				
	\newcommand{\as}{\textcolor{red}{\mathrel{\,:=\,}}}
    \newcommand{\dd}{,\ldots,}
	\newcommand{\dt}[1]{{\em #1}}			
	\newcommand{\wt}[1]{\widetilde{#1}}		
	\providecommand{\ul}[1]{\underline{#1}}	
	\newcommand{\supp}{\mathop{Supp}}
	\newcommand{\set}[1]{\left\{ #1 \right\}}
	\newcommand{\wtb}{\wt{b}}

	\newcommand{\wtf}{{\wt{f}}}
	\newcommand{\wtbfb}{\wt{\bfb}}
	\newcommand{\wtbff}{\wt{\bff}}
	\newcommand{\bfalpha}{{\boldsymbol{\alpha}}}
	
	\newcommand{\bfbeta}{{\boldsymbol{\beta}}}
	\newcommand{\wtbfbeta}{\wt{\boldsymbol{\beta}}}
	\newcommand{\bfb}{{\boldsymbol{b}}}

	\newcommand{\bfdelta}{\boldsymbol{\delta}}
	\newcommand{\bfDelta}{\boldsymbol{\Delta}}
	
	\newcommand{\bfd}{{\boldsymbol{d}}}
	\newcommand{\bff}{{\boldsymbol{f}}}
	
	\newcommand{\bfm}{{\boldsymbol{m}}}

	\newcommand{\bfv}{{\boldsymbol{v}}} 
	\newcommand{\bfz}{{\boldsymbol{z}}} 
	\newcommand{\subx}[1][n]{_{(#1)}} %
	\newcommand{\suby}[1][n-1]{_{(#1)}} 

	\newcommand{\CC}{{\mathbb C}}	
	\newcommand{\DD}{{\mathbb D}}	
	\newcommand{\NN}{{\mathbb N}}	
	\newcommand{\RR}{{\mathbb R}}	
	\newcommand{\ZZ}{{\mathbb Z}}	
	\newcommand{\calO}{{\mathcal O}}	
    
	\renewcommand{\last}[1]{{#1}_n}
    
    \renewcommand{\others}[1]{\mvec{#1}\suby}

	\renewcommand{\spec}[2]{#1(#2)}		
	\newcommand{\beql}[1]{\begin{equation}\label{eq:#1}}
	\newcommand{\eeql}{\end{equation}}
	\newcommand{\beqarray}{\begin{eqnarray}}
	\newcommand{\eeqarray}{\end{eqnarray}}	
	\newcommand{\beqarrays}{\begin{eqnarray*}}
	\newcommand{\eeqarrays}{\end{eqnarray*}} 
	\newcommand{\blem}{\begin{lemma}}
	\newcommand{\elem}{\end{lemma}}
	\newcommand{\bleml}[1]{\begin{lemma} \label{lem:#1}}
	\newcommand{\eleml}{\end{lemma}}
	\newcommand{\blemT}[2]{\begin{lemma}[#1] \label{lem:#2}}
	\newcommand{\elemT}{\end{lemma}}
	\newcommand{\bthm}{\begin{theorem}}
	\newcommand{\ethm}{\end{theorem}}
	\newcommand{\bthml}[1]{\begin{theorem} \label{thm:#1}}
	\newcommand{\ethml}{\end{theorem}}
	\newcommand{\bthmT}[2]{\begin{theorem}[#1] \label{thm:#2}}
	\newcommand{\ethmT}{\end{theorem}}
	\newcommand{\refLem}[1]{Lemma~\protect{\ref{lem:#1}}}
	\newcommand{\refeQ}[1]{(\protect{\ref{eq:#1}})} 
	\newcommand{\refCor}[1]{Corollary~\protect{\ref{cor:#1}}}
	\newcommand{\bitem}{\begin{itemize}}
	\newcommand{\eitem}{\end{itemize}}
	\newcommand{\benum}{\begin{enumerate}}
	\newcommand{\eenum}{\end{enumerate}}
	\newenvironment{pf}[1][Proof.]{{\em #1}}{ 
  	   \hspace*{1mm}\hfill \textbf{ Q.E.D.} \vspace{2mm} \noindent}
	\newcommand{\bpf}[1][Proof.]{\begin{pf}[#1]~} 
	\newcommand{\epf}{\end{pf}}
	\newcommand{\del}[1][z]{\partial_{#1}}

\usepackage{breqn}

\usepackage{mathtools}
\DeclarePairedDelimiter{\ceil}{\lceil}{\rceil}

\begin{document}

\title{Clustering Complex Zeros of Triangular Systems of Polynomials%
  \thanks{R\'emi's work is supported by the European Union's Horizon 2020
    research and innovation programme No. 676541,
	NSF Grants \#~CCF-1563942,
    \#~CCF-1564132 and \#~CCF-1708884.
	Chee's work is supported by NSF Grants
    \#~CCF-1423228 and \#~CCF-1564132.}}
%
\authorrunning{R. Imbach et al.}
%
\author{Remi Imbach\inst{1}
  \and Marc Pouget\inst{2}
  \and Chee Yap\inst{1}} 
     \institute{Courant Institute of Mathematical Sciences, New York
                University, USA\\
                \email{remi.imbach@nyu.edu},
                \email{yap@cs.nyu.edu}
                \and 
                %
         Universite de Lorraine, CNRS, Inria, LORIA, F-54000 Nancy,
                France\\
                \email{marc.pouget@inria.fr} }
\maketitle
 
\begin{abstract}
This paper gives the first algorithm for
finding a set of natural $\epsilon$-clusters
of complex zeros of a regular triangular system 
of polynomials within a given polybox in $\CC^n$,
for any given $\epsilon>0$.
Our algorithm is based on a recent near-optimal algorithm
of Becker et al (2016) for
clustering the complex roots of a univariate polynomial
where the coefficients are represented by number oracles.

\highlightModifs{
Our algorithm is based on recursive subdivision.
It is local, numeric, certified and handles solutions 
with multiplicity.
Our implementation is compared to with well-known
homotopy solvers on various triangular systems.
Our solver always gives
correct answers, is often faster than the homotopy solvers that
often give correct answers, and sometimes faster
than the ones that give sometimes correct results.
}
	 %
	 \keywords{
	 complex root finding
	 \and triangular polynomial system
	 \and near-optimal root isolation
	 \and certified algorithm
	 \and complex root isolation
	 \and oracle multivariable polynomial
	 \and subdivision algorithm
	 \and Pellet's theorem.}	
\end{abstract}
\section{Introduction}
	This
	paper
	considers the fundamental problem of finding the complex
	solutions of a system $\mapf(\vecz)=\veczero$
	of $\dim$ polynomial equations in $\dim$ complex variables
	$\vecz=(z_1,\ldots,z_\dim)$.  The system $\mapf=(f_1\dd
	f_\dim):\C^\dim\rightarrow \C^\dim$ is \dt{triangular} in the sense
	that $f_i\in\C[z_1,\ldots,z_i]$ for $1\leq i \leq \dim$,
	\highlightModifs{where $\deg_{z_i}(f_i)\ge 1$}.
	\highlightModifs{
		As in \cite{boulier2014}, we assume
	that the system is \dt{regular}: this means that
	for each $i$, if $(\alpha_1\dd\alpha_{i-1})$ is a zero of
	$f_{i-1}$ and $c_i(z_1\dd z_{i-1})$ is the leading coefficient
	of $z_i$ in $f_{i}$, then $c_i(\alpha_1\dd \alpha_{i-1})\neq 0$.
	Thus $\mapf$ is a $0$-dimensional system.
	But unlike \cite{boulier2014}, we
	do not assume that the system is square-free:
	indeed the goal of this paper is to demonstrate
	new techniques that can properly
	determine the multiplicity of the root clusters of $\mapf$,
	up to any $\epsilon>0$ resolution.
	}

	Throughout
	this paper, we  use boldface symbols to denote vectors and tuples; for
	instance $\veczero$ stands for $(0,\ldots,0)$.
	
	We are interested in finding clusters of solutions of triangular systems
        and in counting the total multiplicity of solutions in clusters.
        Solving triangular systems is a fundamental task in polynomial equations
        solving, \highlightModifs{since there are many algebraic techniques to
        decompose the original system into triangular systems.}
	
	%
	%
	The problem of isolating the complex solutions of a polynomial system
	in an initial region-of-interest (ROI) is defined as follows: let
	$\solsIn{\vecB}{\mapf}$ denote the set of solutions of $\mapf$ in
	$\vecB$, regarded\footnote{ A \dt{multiset} $S$ is a pair
	$(\ul{S},\mu)$ where $\ul{S}$ is an ordinary set called the
	\dt{underlying set} and $\mu:\ul{S}\to\N$ assigns a positive integer
	$\mu(x)$ to each $x\in \ul{S}$.  Call $\mu(x)$ the \dt{multiplicity}
	of $x$ in $S$, and $\mu(S)\as\sum_{x\in\ul{S}}\mu(x)$ the \dt{total
	multiplicity} of $S$.  Also, let $|S|$ denote the cardinality of
	$\ul{S}$.  If $|S|=1$, then $S$ is called a \dt{singleton}.   We can
	form the union $S\cup S'$ of two multisets with underlying set
	$\ul{S}\cup\ul{S'}$, and the multiplicities add up as expected.  } as
	a multiset.
		
	\begin{center} \fbox{ \begin{minipage}{0.9 \textwidth} \noindent
		\textbf{Local Isolation Problem (LIP):}\\ \noindent
		\textbf{Given:} a polynomial map $\mapf:\C^\dim\rightarrow
		\C^\dim$, a polybox $\vecB\subset\C^\dim$, $\epsilon>0$\\
		\noindent \textbf{Output:} a set $\{\vecD^1,\ldots,\vecD^l\}$ of
		pairwise disjoint polydiscs of radius $\leq\epsilon$ where\\
		\noindent \hphantom{\textbf{Output:}} - $\solsIn{\vecB}{\mapf} =
		\bigcup_{j=1}^{l} \solsIn{\vecD^j}{\mapf}$.\\ \noindent
	\hphantom{\textbf{Output:}} - each $\solsIn{\vecD^j}{\mapf}$ is a
	singleton.  \end{minipage} } \end{center}
	
	This is ``local'' because we restrict attention to roots in a ROI
	$\vecB$.  There are two issues with (LIP) as formulated above:
	deciding if $\solsIn{\vecD^j}{\mapf}$ is a singleton, and deciding if
	such a singleton lies in $\vecB$, are two ``zero problems'' that
	require exact computation.  Generally, this can only be decided if
	$\mapf$ is algebraic.  Even in the algebraic case, this may be very
	expensive.  In \cite{yap-sagraloff-sharma:cluster:13,2016Becker} these
	two issues are side-stepped by defining the local clustering problem
	which is described next.  
	
	Before proceeding, we fix some general notations for this paper.  A
	\emph{polydisc} $\vecD$ is a vector $(\Delta_1,\ldots,\Delta_\dim)$ of
	complex discs.  The \emph{center} of $\vecD$ is the vector of the
	centers of its components and the \emph{radius} $\radius{\vecD}$ 
	of $\vecD$ 
	is the vector of the radii 
	of its components.  If $\delta$ is any positive real number, we denote
	by $\delta\vecD$ the polydisc
	$(\delta\Delta_1,\ldots,\delta\Delta_\dim)$ that has the same center
	as $\vecD$ and radius $\delta \radius{\vecD}$.
	We also say $\radius{\vecD}\le\delta$ if each component of
	$\radius{\vecD}$ is $\le\delta$.  A \emph{(square complex) box} $B$ is
	a complex interval $[\ell_1,u_1]+\ii([\ell_2,u_2])$ where
	$u_2-\ell_2=u_1-\ell_1$ and $\ii\as \sqrt{-1}$; the \emph{width}
	$\width{B}$ of $B$ is $u_1-\ell_1$ and the \emph{center} of $B$ is
	$u_1+\frac{\width{B}}{2} + \ii(u_2+\frac{\width{B}}{2})$.  A
	\emph{polybox}
	$\vecB\ib\C^\dim$
	is the set $\prod_{i=1}^n B_i$ which is represented by the
		vector $(B_1,\ldots,B_\dim)$ of boxes.
	The \emph{center} of $\vecB$  is the vector of the centers of its
	components; the \emph{width} $\width{\vecB}$ of $\vecB$ is the max of
	the widths of its components.
	If $\delta$ is any positive real number, we denote by $\delta\vecB$
	the polybox $(\delta B_1,\ldots,\delta B_\dim)$ that has the same
	center than $\vecB$ and width $\delta \width{\vecB}$.
	It is also convenient to identify $\vecD$ as the subset
	$\prod_{i=1}^\dim \vecD_i$ of $\CC^n$; a similar remark applies
	to $\vecB$.
	
	We introduce three notions to define the local solution clustering
	problem.  Let $\veca\in\C^\dim$ be a solution of
	$\mapf(\vecz)=\veczero$. 
	The \emph{multiplicity} of $\veca$ in $\mapf$, also called the
	\emph{intersection multiplicity} of $\veca$ in $\mapf$ is classically
	defined by localization of rings as in \cite[Def.~1,
	p.~61]{zhang2011real}, we denote it by $\multipli{\veca}{\mapf}$.  An
	equivalent definition uses dual spaces, see \cite[Def.~ 1,
	p.~117]{dayton2005computing}.
	For any set $S\subseteq\C^\dim$, we denote by $\solsIn{S}{\mapf}$ the
	multiset of zeros (i.e., solutions) of $\mapf$ in $S$, and
	$\nbSolsIn{S}{\mapf}$ the total multiplicity of $\solsIn{S}{\mapf}$.
	If $S$ is a polydisc, we
	call $\solsIn{S}{\mapf}$ a \dt{cluster} if it is non-empty, and $S$ is
	an \dt{isolator} of the cluster.  If in addition, we have that
	$\solsIn{S}{\mapf}=\solsIn{3\cdot S}{\mapf}$, we call
	$\solsIn{S}{\mapf}$ a \dt{natural cluster} and call $S$ a \dt{natural
	isolator}.
	In the context of numerical algorithm, the notion of cluster of
	solutions is more meaningful than that of solution with multiplicity
	since the perturbation of a multiple solution generates a cluster.  We
	thus ``soften'' the problem of isolating the solutions of a triangular
	system of polynomial equations while counting their multiplicities by
	translating it into the local solution clustering problem defined as
	follows:
	\begin{center} \fbox{
		\begin{minipage}{0.9 \textwidth} \noindent
		\textbf{Local Clustering Problem (LCP):}\\ \noindent
		\textbf{Given:} a polynomial map $\mapf:\C^\dim\rightarrow
		\C^\dim$, a polybox $\vecB\subset\C^\dim$, $\epsilon>0$\\
		\noindent \textbf{Output:} a set of pairs
		$\{(\vecD^1,m^1),\ldots,(\vecD^l,m^l)\}$ where: \\ \noindent
		\hphantom{\textbf{Output:}} - the $\vecD^j$s are pairwise disjoint
		polydiscs of radius $\leq\epsilon$,\\ \noindent
		\hphantom{\textbf{Output:}} - each $m^j =
		\nbSolsIn{\vecD^j}{\mapf}=\nbSolsIn{3\vecD^j}{\mapf}$ \\ \noindent
		\hphantom{\textbf{Output:}} - $\solsIn{\vecB}{\mapf} \subseteq
		\bigcup_{j=1}^{l} \solsIn{\vecD^j}{\mapf} \subseteq
		\solsIn{2\vecB}{\mapf}$.
	\end{minipage}
	} \end{center}
	In this (LCP) reformulation of (LIP), we have removed the two ``zero
	problems'' noted above: we output clusters to avoid the first problem,
	and we allow the output to contain zeroes outside the ROI $\vecB$ to
	avoid the second one.  We choose $2\vecB$ for simplicity; it is easy
	to replace the factor of $2$ by $1+\delta$ for any desired $\delta>0$.
	
	\paragraph{Overview.} In the remaining of this section we explain our
	contribution, summarize previous work and the local univariate
	 clustering method of \cite{2016Becker}. In
	Sec.~\ref{section_multiplicities}, we define the notion of \emph{tower
	of clusters} together with a recursive method to compute the sum of
	multiplicities of the solutions it contains. 
        Sec.~\ref{section_suffCond} analyzes the loss of precision induced by
        approximate specialization.
        Our algorithm for solving
	the local  clustering problem for triangular systems is
	introduced in Sec.~\ref{section_clustering}. The implementation and
	experimental results are presented in Sec.~\ref{sec_benchmark}.
	
	\subsection{Our contributions}
	We propose an algorithm for solving the complex  clustering
	problem for a triangular system $\mapf(\vecz)=\veczero$ with a
	zero-dimensional solution set.  To this end, we propose a formula to
	count the sum of multiplicities of solutions in a cluster.
	Our formula is derived from a result of \cite{zhang2011real} that
	links the intersection multiplicity of a solution of a triangular
	system to multiplicities in fibers.
	We define  \emph{towers of clusters} to encode clusters of solutions
	of a triangular system in stacks (or towers) of clusters of roots of
	univariate polynomials.
	
	Our algorithm exploits the triangular form of $\mapf=(f_1\dd f_n)$:
	the standard idea is to recursively find roots
	of the form $(\alpha_1\dd\alpha_{\dim-1})$ of 
	$f_1=\cdots=f_{\dim-1}=0$, then substituting them into
	$f_\dim$ to obtain a univariate polynomial
	$g_n(z_n)=f_n(\alpha_1\dd \alpha_{\dim-1}, z_n)$.
	If $\alpha_n$ is a root of $g_n(z_n)$, then we
	have have extended the solution to $(\alpha_1\dd \alpha_n)$
	of the original $\mapf$.
	The challenge is to extend this idea to compute clusters
	of zeros of $\mapf$ from clusters of
	zeros of $f_1=\cdots=f_{\dim-1}=0$.
	Moreover, we want to allow
	the coefficients of each $f_i$ to be oracle numbers.
	The use of oracle numbers allows us to treat polynomial
	systems whose coefficients are algebraic numbers and beyond.
	
	To compute clusters of roots of a univariate polynomial given as an
	oracle, we rely on the recent algorithm described in
	\cite{2016Becker},
	based on a predicate introduced in \cite{BECKER2017} that combines
	Pellet's theorem and Graeffe iterations to determine the number of
	roots counted with multiplicities in a complex disc; this predicate is
	called \emph{soft} because it only requires the polynomial to be known
	as approximations. It is used in a subdivision framework combined with
	Newton iterations to achieve a near optimal complexity.

	\highlightModifs{
	We implemented our algorithm and made it available 
	as 
        the \Julialang\footnote{\url{
	https://julialang.org/}} 
	package \CCJL\footnote{
	\url{
	https://github.com/rimbach/Ccluster.jl}}.
	Our experiments show that it advantageously compares 
	 to major homotopy solvers 
	for solving random dense triangular systems 
	in terms of solving times and reliability
	(\emph{i.e.} getting the correct number of solutions
	and the correct multiplicity structures).
	Homotopy solving is more general because
	it deals with any polynomial system.
	We also propose experiments with
	triangular systems obtained with elimination
	procedures.
	}

\subsection{Related work}
\label{subsection_biblio}
	There is a vast literature on solving polynomial systems and we can only refer
	to book surveys and references therein, see for instance
	\cite{2005EmirisDickenstein,wampler2005numerical}. On the algebraic side,
	symbolic tools like Groebner basis, resultant, rational univariate
	parametrization or triangularization, 
	\highlightModifs{
	find an equivalent triangular system or set
	thus reducing the problem to the univariate case. }
	Being symbolic,
	these methods handle all input, in particular with solutions with
	multiplicities, and are certified but at the price of a high complexity that
	limits their use in practice.
	\highlightModifs{Implementations of hybrid symbolic-numeric solvers are available
	for instance in {\tt Singular}\footnote{\url{https://www.singular.uni-kl.de/}} 
        via {\tt solve.lib}
	or in \maple via 
        {\tt RootFinding[Isolate]}.}
	
	On the numerical side, one can find subdivision and homotopy methods.  The main
	advantage of subdivision methods is their locality: the practical complexity
	depends on the size of the solving domain and the number of solutions in this
	domain. Their main drawback is that they are only practical for low dimensional
	systems. On the other hand, homotopy methods are efficient for high dimensional
	systems, they are not local but solutions are computed independently from one
	another.  
	Numerical methods only work for restricted classes of systems and the
	certification of the output remains a challenge. Multiprecision arithmetic, interval
	analysis, deflation and $\alpha$-theory are now classical tools to address this 
	certification issue \cite{Giusti2007,MKCbook09,beltran,Xu2018approach}.

	In the univariate case, practical certified algorithms are now available for
	real and complex solving that match the best known complexity bounds together
	with efficient implementations \cite{Kobel,ICMSpaper}.
	\highlightModifs{
	For the bivariate case, the problem of solving a triangular system can be seen
	as a univariate isolation in an extension field. The most recent contributions
	in this direction presenting algorithms together with complexity analysis are
	\cite{2018arXiv180710622N,STRZEBONSKI2019}.}

	Only a few work address the specific problem of solving triangular polynomial
	systems.  The solving can then be performed coordinate by coordinate by
	specialization and univariate solving in fibers.  When the systems only have
	regular solutions, extensions of classical univariate isolation algorithms to
	polynomial with interval coefficients have been proposed
	\cite{Collins02,ekkmsw-casc-2005,boulier2014}. In the presence of multiple
	solutions, one approach is to use a symbolic preprocessing to further decompose
	the system in regular sub-systems. Another approach is the sleeve method with
	separation bounds \cite{cheng2009complete}.  The authors of \cite{zhang2011real}
	propose a formula to compute the multiplicity of a solution of a triangular
	system: the latter multiplicity is the product of the multiplicities of the
	components of a solution in the fibers. Then, by using square free factorization
	of univariate polynomials specialized in fibers, they describe an algorithm to
	retrieve the real solutions of a triangular system with their multiplicities.
	\highlightModifs{
	In \cite{Li-CASC-2012}, the method of Local Generic Position is adapted to the
	special case of triangular systems with the advantage of only using resultant
	computations (instead of Goebner basis), multiplicities are also computed.}

\subsection{Definitions and Notation}
\label{subsection_definitions}

{\bf Convention for Vectors.}
	We introduce some general conventions for vectors that will
	simplify the following development.
	Vectors are indicated by bold fonts.  If $\bfv=(v_1\dd v_n)$
	is an $n$-vector, and $i=1\dd n$, then the $i$-th component
	$v_i$ is\footnote{
			In general, $\bfv_i\neq v_i$ since $\bfv$ and $v_i$
			are independent variables.  So our bold font variables $\bfv$
			do not entail the existence of non-bold font counterparts
			such as $v_i$.
	}
	denoted $\bfv_i$ and
	the $i$-vector $(v_1\dd v_i)$ is denoted $\bfv\subx[i]$.
	Thus $\bfv=(\bfv\subx[n-1],\bfv_n)$, and
	``$\bfv=\bfv\subx[n]$'' is an idiomatic way
	of saying that $\bfv$ is an $n$-vector.
	Because of the subscript convention, we
	will superscripts such as $\bfv^1, \bfv^2$, etc, to distinguish
	among a set of related $n$-vectors.

{\bf Normed Vector Spaces.} 
	In order to do error analysis, we need to treat
	$\CC[\bfz]$ and $\CC^n$ as normed vector spaces:
	for $f\in\CC[\bfz]$ and $\bfb\in\CC^n$,
	let
	$\|f\|$ and $\|\bfb\|$ denote the infinity norm
	on polynomials and vectors, respectively.
	We use the following \dt{perturbation convention}:
	let $\delta\ge 0$.  Then we will write
	$f\pm \delta$ to denote some polynomial
	$\wtf\in \CC[\bfz]$ that satisfies $\|f-\wtf\|\le \delta$.
	Similarly, $\bfb\pm \delta$ denotes some vector $\wtbfb\in\CC^n$
	that satisfies $\|\bfb-\wtbfb\|\le \delta$.
	If $\delta\le 2^{-L}$ then $\wtbfb$ and $\wtbff$
	are called \dt{$L$-bit approximations}
	of $\bfb$ and $\bff$, respectively.
	
	We define the \dt{degree sequence} of $f\in\CC[\bfz]$
	to be $\bfd=\bfd(f)$ where $\bfd_i$ is the degree of $z_i$ in $f$.
	If $\bfb\in \CC^k$ ($k=1\dd n$), let
	$f(\bfb)$ denote the polynomial that results
	from the substitution 
	$\bfz_i \to \bfb_i$ (for $i=1\dd k$).
	The result is a polynomial
	$f(\bfb)\in \CC[z_{k+1}\dd z_n]$ called the
	\dt{specialization} of $f$ by $\bfb$.
	Note that $f(\bfb)$ is a polynomial in at most $n-k$ variables.
	In particular, when $n=k$, then $f(\bfb)$ is a constant
	(called the \dt{evaluation} of $f$ at $\bfb$).
	For instance, suppose $\bfb\in\CC^n$,
	then $f(\bfb\subx[n-1])$ is a polynomial in $z_n$
	and $f(\bfb\subx[n-1])(\bfb_n)=f(\bfb)$.

\ignore{
	For any $\dim$-dimensional vector $\mvec{v}=(v_1,\ldots,v_\dim)$
	with $\dim>1$,
	we  denote by $\last{v}$ its last component
	(\emph{i.e.} $v_\dim$)
	and $\others{v}$
	for the $(\dim-1)$-dimensional vector $(v_1,\ldots,v_{\dim-1})$
	of its other components.
	By abuse of notation, we  sometimes write 
	$\mvec{v}=(\others{v},\last{v})$. The modulus of a complex number $z$
	is noted
	$|z|$ and the norm of a vector is $\|v\| = \max_{i=1\dots n} |v_i|$.
	
	If $\vecb$ is any point in $\C^{\dim-1}$,
	$\vecz=(z_1,\ldots,z_\dim)$
	and $g\in\C[\vecz]$,
	we denote by 
	$\spec{g}{\vecb}$
	the univariate polynomial $g(\vecb,\last{\vecz})$ 
	in $\C[\last{\vecz}]$ 
	obtained by specializing $g$ at $\vecb$.
	If $\veca\in\C^{\dim}$ is a solution of $\mapf$,
	we call multiplicity of $\last{\veca}$ in $\last{\mapf}$ 
	in the fiber $\others{\veca}$
	the multiplicity of the root $\last{\veca}$ of the 
	polynomial $(\last{\mapf})_{\others{\veca}}$.
	We denote it by
	$\multipli{\last{\veca}}{(\last{\mapf})_{\others{\veca}}}$.
	}%
    
If $B\ib\C$ is a box
with center $c$ and width $w$,
we denote by $\contDisc{B}$ the disc with center $c$ and 
radius $\frac{3}{4}w$. Note that $\contDisc{B}$ contains $B$.
If $\vecB\subset\C^\dim$ is a polybox,
let $\contDisc{\vecB}$ be the polydisc 
where $\contDisc{\vecB}_i = \contDisc{\vecB_i}$.
    
\ignore{
	A polydisk $\vecD$ is called an \emph{isolator} if 
	$\nbSolsIn{\vecD}{\mapf} = \nbSolsIn{3\vecD}{\mapf}$.
	Any non-empty set of the form $\solsIn{\vecD}{\mapf}$
	is called a \emph{cluster}
	of solutions of $\mapf(\vecz)=\veczero$, 
	and it is a \emph{natural} cluster if $\vecD$
	is an isolator.
}%

\ignore{%
	We call $L$-bit approximation of $a\in\C$ a dyadic complex number
	$\tilde{a}$ that satisfies $|a-\tilde{a}|\leq 2^{-L}$.  We call $L$-bit
	approximation of $\veca=(a_1,\ldots,a_\dim)\in\C^\dim$ a vector
	$\tilde{\veca}\in\C^\dim$ such that $\tilde{a_i}$ is an $L$-bit
	approximation of $a_i$ for $1\leq i\leq \dim$, in other words
	$||\tilde{\veca} -\veca||\leq 2^{-L}$.  If $g$ is a univariate
	polynomial, we call $L$-bit approximation of $g$ a univariate
	polynomial $\tilde{g}$ which coefficients are $L$-bit approximations of
	the coefficients of $g$.
	}%
	
{\bf Oracle Computational Model.} 
	We use two kinds of numbers in our algorithms:
	an explicit kind which is standard in computing,
	and an implicit kind which we call ``oracles''.
	Our explicit numbers are dyadic numbers (i.e., bigFloats),
	$\DD\as \set{n2^m: n,m\in\ZZ}$.  A pair $(n,m)$
	of integers represents the \dt{nominal value} of $n2^m\in \DD$.
	However, we also want this pair to represent the interval
	$[(n-\half)2^m, (n+\half)2^m]$.  To distinguish between them,
	we write $(n,m)_0$ for the nominal value, and $(n,m)_1$ 
	for the interval of width $2^m$.
	Call $(n,m)_1$ an \dt{$L$-bit dyadic interval}
	if $m\le -L$ (so the interval has width at most $2^{-L}$).
	Note that $(2n,m)_1$ and $(n,m+1)_1$ are different despite having the
	same nominal value.  As another example, note that $(0,m)_1$
	is the interval $[-2^{m-1},2^{m-1}]$.
	When we say a box, disc, polybox, etc, is \dt{dyadic}, it means that
	all its parameters are given by dyadic numbers.
	The set of closed intervals with dyadic endpoints is
	denoted $\intbox \DD$.  Also, let $\intbox^n\DD$ denote
	$n$-dimensional dyadic boxes.
	
	The implicit numbers in our algorithms are functions: for any
	real number $x\in\RR$, an \dt{oracle for $x$} is a function
	$\calO:\ZZ\to\intbox\DD$ such that $\calO_x(L)$ is an $L$-bit
	dyadic interval containing $x$.  There is normally no confusion in
	identifying the real number $x$ with {\em any}
	oracle function $\calO_x$ for $x$.  Moreover, we write $(x)_L$
	instead of $\calO_x(L)$.
	E.g., if $x$ is a real algebraic number with defining polynomial
	$p\in \ZZ[X]$ and isolating interval $I$, we may define
	an oracle $\calO_x=\calO(p,I)$ for $x$ in a fairly standard way.
	Next, an oracle $\calO_z$ for a complex number $z=x+\ii y$
	is a function $\calO_z:\ZZ\to \intbox^2\DD$ such that
	$\calO_z(L)=\calO_x(L)+\ii\calO_y(L)$ where 
	$\calO_x,\calO_y$ are oracles for $x$ and $y$.   Again, we may
	identify $z$ with any oracle $\calO_z$, and write $(z)_L$
	instead of $\calO_z(L)$.  For polynomials $f\in\CC[\bfz\subx]$
	in $n\ge 2$ variables, we assume a sparse representation,
		$f=\sum_{\bfalpha\in \supp(f)} f_{\bfalpha} \bfz^\bfalpha$
	with fixed support $\supp(f)\ib\NN^n$,
	with coefficients $f_\bfalpha\in\CC\setminus\set{0}$, and
	$\bfz^\bfalpha\as \prod_{i=1}^n \bfz_i^{\bfalpha_i}$
	are power products.
	An oracle $\calO_f$ for $f$ amounts to having oracles for
	each coefficient $f_\bfalpha$ of $f$.  Moreover $\calO_f(L)$
	may be written $(f)_L$
	is the interval polynomial whose coefficients are $(f_\alpha)_L$.
	Call $(f)_L$ a \dt{dyadic interval polynomial}.

\ignore{
	We approximate real numbers with dyadic numbers (i.e., bigFloats),
	$\ZZ[\half]=\set{n2^m: n,m\in\ZZ}$.
	Our computational model views
	real numbers $x\in\RR$ as oracles:
	given any integer $L$, the oracle $x$ returns an $L$-bit
	dyadic number $(x)_L$ that is an $L$-bit approximation of $x$.
	It is an absolute (not relative) approximation:
	$|x-(x)_L|\le 2^{-L}$.
	This extends to complex numbers: the oracle for $z\in\CC$
	returns $(z)_L$, a pair of dyadic numbers
	approximating the real and complex part of $z$ to $L$-bits.
	For polynomials in $f\in\CC[\bfz]$,
	we assume a sparse representation,
		$f=\sum_{\bfalpha\in \supp(f)} f_{\bfalpha} \bfz^\bfalpha$
	where $\supp(f)\ib\NN^n$ is fixed, 
	$f_\bfalpha\in\CC\setminus\set{0}$ (non-zero coefficients),
	and 
	$\bfz^\bfalpha\as \prod_{i=1}^n \bfz_i^{\bfalpha_i}$
	are power products.
	The oracle for $f$ produces, for an integer $L$, a
	polynomial $(f)_L$ whose support $\supp((f)_L)$ is
	a subset of $\supp(f)$ such that each coefficient
	of $(f)_L$ is an $L$-bit approximation of the corresponding
	coefficient of $f$.  Clearly,
	if $\bfalpha\in \supp(f)$ does not appear in
	$\supp((f)_L)$, it is because $|f_\bfalpha|<2^{-L}$.
	In some applications, we may need 
	to ensure that $L$ is large enough so that the ``leading''
	coefficient (however defined) does not vanish.
}
\subsection{Oracles for Root Cluster of Univariate Polynomials}
\label{subsection_Pellet}
	The starting point for this paper is the fundamental result
	that the Local Clustering Problem (LCP) has been
	solved in the univariate setting:
	
	\bproT{See \cite{2016Becker,BECKER2017}}{unisolve}
	There is an algorithm $Cluster(f, B,\epsilon)$
	that solves the Local Clustering Problem
	when $f:\CC\to\CC$ is a univariate oracle polynomial.
	\eproT
	In other words, the output of 
	$Cluster(f, B,\epsilon)$ is a set $\set{(\Delta_i,m_i): i=1\dd k}$
	such that each $\Delta_i$ is a natural $\epsilon$-isolator, and
			
			$$\solsIn{B}{f} \ib \bigcup_{i=1}^{k} 
				\solsIn{\Delta_i}{f}
				\subseteq \solsIn{2B}{f}.$$

	To make this result the basis of our multivariate
	clustering algorithm, we need to generalize this result.
	In particular, we need to be able to further
	refine each output $(\Delta_i,m_i)$ of this algorithm.
	If $(\Delta_i,m_i)$ represents the cluster $C_i$ of roots,
	we want to get better approximation of $C_i$, i.e.,
	we want to treat $C_i$ like number oracles.
	Fortunately, the algorithm in \cite{2016Becker,BECKER2017}
	already contains the tools to do this.  What is lacking is 
	a conceptual framework to capture this.

	Our goal is to extend the concept of number oracles
	to ``cluster oracles''.  To support the several modifications
	which are needed, we revise our previous view of ``oracles as
	functions''.  We now think of an oracle
	$\calO$ as a computational object with \dt{state information},
	and which can transform itself in order to
	update its state information.
	For any $L\in\ZZ$, recall that $\calO(L)$ is
	a dyadic object that is at least $L$-bit accurate.
	E.g., if $\calO$ is the oracle for $x\in\RR$,
	$\calO(L)$ is an interval containing $x$ of width $\le 2^{-L}$.
	But now, we say that oracle is transformed to a
	new oracle which we shall denote by ``$(\calO)_L$''
	whose state information is $\calO(L)$.  In general,
	let $\sigma(\calO)$ denote the state information in $\calO$.
	Next,
	for a cluster $C\ib\CC$ of roots of a univariable polynomial
	$p(z)\in\CC[z]$, its oracle $\calO_C$ has state $\sigma(\calO_C)$
	that is a pair $(\Delta, m)$ where $\Delta\ib\CC$ is a dyadic disc
	satisfying $C = \solsIn{\Delta}{p}=\solsIn{3\Delta}{p}$
	and $m$ is the total multiplicity of $C$.
	Thus $C$ is automatically a natural cluster.
	We say $\calO_C$ is \dt{$L$-bit accurate} if the radius
	of $\Delta$ is at most $2^{-L}$.
	Intuitively, we expect $(\calO_C)_L$ to be
	an oracle for $C$ that is $L$-bit accurate.
	Unfortunately, this may be impossible unless $C$ is a singleton
	cluster.  In general, we may have to split $C$ into two
	or more clusters.  We therefore need one more extension:
	the map $L\mapsto (\calO_C)_L$ returns a set
		
		$$\set{\calO_{C_1}\dd \calO_{C_k}},
				\quad (\textrm{for some } k\ge 1)$$
	of cluster oracles with the property that
	$C=\cup_{i=1}^k C_i$ (union of multisets), and
	each $\calO_{C_i}$  is $L$-bit accurate.
	We generalize \refPro{unisolve} so that it outputs
	a collection of cluster oracles:

		\bproT{See \cite{2016Becker,BECKER2017,ICMSpaper}}{clussolve}
		\label{prop_clussolve}
		Let $\calO_f$ be an oracle for a univariate polynomial
		$f:\CC\to\CC$.
		There is an algorithm $ClusterOracle(\calO_f, B, L)$
		that returns a set $\set{\calO_{C_i}: i=1\dd k}$
		of cluster oracles such that
			
			$$\solsIn{B}{f}
				\ib \bigcup_{i=1}^k C_i \ib \solsIn{2B}{f}.$$
		and each $\calO_{C_i}$ is $L$-bit accurate.
		\eproT
		
\ignore{
	We distinguish between a solution cluster and its representation.
	A pair $(\vecD,m)$ is called a 
	\dt{cluster representation} (relative to $\mapf$) if
	$\vecD\ib\C^\dim$ is a polydisc satisfying $m=\#(\vecD,\mapf)$.
	We can also call $(\vecD,m)$ a \dt{representation of the
	cluster} $\solsIn{\vecD}{\mapf}$.
	All representations will be relative to some $\mapf$;
	we may omit $\mapf$ when it is understood.
	If $(\vecD,m)$ and $(3\vecD,m)$ are both cluster representations,
	we call $(\vecD,m)$ a \dt{natural representation}
	(of the natural cluster $\solsIn{\vecD}{\mapf}$).
}%

\ignore{
	We recall in Algo~\ref{TstarLtest} the specifications of the
	\TsTL~test introduced in \cite{BECKER2017}.
	It combines Pellet's theorem
	and Graeffe iterations to determine the number of solutions 
	counted with multiplicities of a polynomial $f\in\C[z]$ in a 
	complex disc $\Delta$, using an $L$-bit approximation $\app{f}$
	of $f$. It returns $-2$ if $L$ is to small to decide,
	$-1$ if $f$ has roots close to the boundary of $\Delta$
	and an integer $k\geq 0$ only if $f$ has $k$ roots counted
	with multiplicities in $\Delta$.
	
	}
	
\section{Sum of multiplicities in clusters of solutions}
\label{section_multiplicities}
We extend in Sec.~\ref{subsection_multiplicities} a result of \cite{zhang2011real} to
an inductive formula giving the sum of multiplicities of solutions of a
triangular system in a cluster.  
In Sec.~\ref{subsection_TAC}, we introduce 
a representation of clusters of solutions of $\mapf$ 
called \emph{tower representation}, reflecting the triangular form of $\mapf$.
Sec.~\ref{subsection_examples} presents two illustrative examples.

\subsection{Two examples}
\label{subsection_examples}

\begin{figure}[t]
 {\centering
  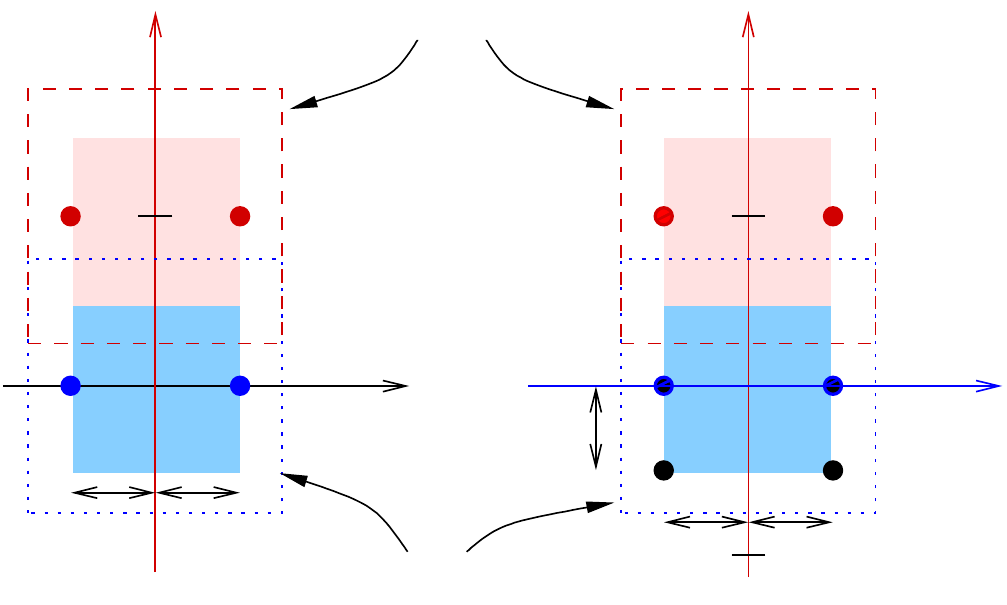\\}
 \fbox{
\begin{minipage}{0.95\textwidth}
$
 \begin{array}{ll}
  \multipli{\veca^3}{\mapg}=1\times 1 & \multipli{\veca^2}{\mapg}=1\times 1\\
  \multipli{\veca^4}{\mapg}=1\times 1 & \multipli{\veca^1}{\mapg}=1\times 1 \\
  \nbSolsIn{\vecD(\vecB^1)}{\mapg}=2\times 1 & \\
  \nbSolsIn{\vecD(\vecB^2)}{\mapg}=2\times1  & \\
 \end{array} \hspace{0.4cm}
 \begin{array}{ll}
  \multipli{\veca^3}{\maph}=1\times 1 & \multipli{\veca^2}{\maph}=2\times 1 \\
  \multipli{\veca^4}{\maph}=1\times 1 & \multipli{\veca^1}{\maph}=2\times 1 \\
  \multipli{\veca^5}{\maph}=1\times 2 & \multipli{\veca^6}{\maph}=2\times 2 \\
  \nbSolsIn{\vecD(\vecB^1)}{\maph}=3\times 3 & \\
  \nbSolsIn{\vecD(\vecB^2)}{\maph}=3\times1  &\\
 \end{array}\\
$
\end{minipage}
}
 \caption{
 On the left (resp. right), the solutions of $\mapg(\vecz)=0$ (resp. $\maph(\vecz)=0$)
 defined in Eq.~\ref{eq_syst_1} (resp. Eq.~\ref{eq_syst_2}) with $\delta = 1$.
 $\vecB^1$ (resp. $\vecB^2$) is the polybox of $\C^2$
 with center $(0,0)$ (resp. $(0,1)$) and width $2*2^{-\delta}$.
 The boxes in dashed lines are the real parts of $\vecD(\vecB^1)$
 and $\vecD(\vecB^2)$.
 In the frame, the multiplicities of solutions of each system
are  computed with the formula of Zhang (\refPro{zhang}) and
Thm.~\ref{th_count}.} 
 \label{fig_sols}
\end{figure}

Let $\delta>0$ be an integer.
We define the systems
$\mapg(\vecz)=(g_1(z_1),g_2(z_1,z_2))=\veczero$ and 
$\maph(\vecz)=(h_1(z_1),h_2(z_1,z_2))=\veczero$ 
as follows:

\begin{equation}
 \label{eq_syst_1}
 (\mapg(\vecz)=\veczero):
 \left\{ \begin{array}{rcl}
         (z_1-2^{-\delta})(z_1+2^{-\delta}) & = & 0\\
         (z_2-2^{2\delta}z_1^2)z_2          & = & 0\\
        \end{array}\right.
\end{equation}

\begin{equation}
 \label{eq_syst_2}
 (\maph(\vecz)=\veczero):
 \left\{ \begin{array}{rcl}
         (z_1-2^{-\delta})^2(z_1+2^{-\delta}) & = & 0\\
         (z_2+2^{\delta}z_1^2)^2(z_2-1)z_2 & = & 0\\
        \end{array}\right.
\end{equation}
$\mapg(\vecz)=0$ has 4 solutions:
$\veca^1=(2^{-\delta},0)$,
$\veca^2=(2^{-\delta},1)$,
$\veca^3=(-2^{-\delta},1)$ and
$\veca^4=(-2^{-\delta},0)$.
$\maph(\vecz)=0$ has 6 solutions:
$\veca^1$,
$\veca^2$,
$\veca^3$,
$\veca^4$,
$\veca^5=(-2^{-\delta},-2^{-\delta})$ and 
$\veca^6=(2^{-\delta},-2^{-\delta})$.
For $1\leq i\leq 6$, let $\veca^i=(a_1^i,a_2^i)$.
The solutions of both $\mapg=0$ and $\maph=0$
are depicted in Fig.~\ref{fig_sols}.

\subsection{Sum of multiplicities in a cluster}
\label{subsection_multiplicities}

We recall a theorem of Zhang \cite{zhang2011real} 
for counting multiplicities of solutions of triangular systems,
based multiplicities in fibers.  We may rephrase it inductively:
	
\bproT{\cite{zhang2011real}}{zhang}
	\label{th_sols}
	Let $\dim\ge 2$ and $\veca\in\C^{\dim}$ be a solution of 
	the triangular system $\mapf(\vecz)=0$.
	The multiplicity of $\veca$ in $\mapf$ is
	
	\[
	\multipli{\veca}{\mapf} = 
	\multipli{\last{\veca}}{\spec{\last{\mapf}}{\others{\veca}}}
	\times
	\multipli{\others{\veca}}{\others{\mapf}}.
	\]
\eproT

We extend \refPro{zhang} to a formula giving the 
total multiplicity of a cluster $\solsIn{\vecD}{\mapf}$.

\begin{theorem}
\label{th_count}
Let $\solsIn{\vecD}{\mapf}$ be a cluster of solutions
of the triangular system $\mapf(\vecz)=0$.
If there is an integer $m\geq 1$ so that for any solution
$\veca\in\solsIn{\vecD}{\mapf}$, one has  
$m=\nbSolsIn{\last{\vecD}}{\spec{\last{\mapf}}{\others{\veca}}}$,
then 

\[
 \nbSolsIn{\vecD}{\mapf}=m\times\nbSolsIn{\others{\vecD}}{\others{\mapf}}.
\]
where $\nbSolsIn{\others{\vecD}}{\others{\mapf}}=1$ when $\dim=1$.
\end{theorem}

%
Let us apply \refPro{zhang}
to compute the multiplicities of solutions
of $\mapg(\vecz)=0$ and $\maph(\vecz)=0$
(see Eq.~\ref{eq_syst_1} and Eq.~\ref{eq_syst_2}).
$\veca^1$ has multiplicity $1$ in $\mapg$: 
$\multipli{\veca^1}{\mapg}=\multipli{a_1^1}{g_1}\times\multipli{a_2^1}{g_2({a_1^1)}}=1\times1$.
$\veca^1$ has multiplicity $2$ in $\maph$: 
$\multipli{\veca^1}{\maph}=\multipli{a_1^1}{h_1}\times\multipli{a_2^1}{h_2({a_1^1})}=2\times1$.
The multiplicities of other solutions are
given in fig.~\ref{fig_sols}.

Let $\vecB^1=(B^1_1,B^1_2)$
be the polybox centered
in $(0,0)$ having width $2\times2^{-\delta}$.
$\solsIn{\vecD(\vecB^1)}{\mapg}
=\{\veca^1,\veca^4\}$ and
$\nbSolsIn{\vecD(\vecB^1)}{\mapg}=2$.
Since $\nbSolsIn{\contDisc{B^1_2}}{\spec{\last{\mapg}}{\others{\veca^1}}}
= \nbSolsIn{\contDisc{B^1_2}}{\spec{\last{\mapg}}{\others{\veca^4}}}=1$,
applying Thm.~\ref{th_count}
yields $\nbSolsIn{\vecD(\vecB^1)}{\mapg}=2\times 1$.


$\solsIn{\vecD(\vecB^1)}{\maph}
=\{\veca^1,\veca^4, \veca^5, \veca^6\}$ and
$\nbSolsIn{\vecD(\vecB^1)}{\maph}=9$.
Again, one has \\
$\nbSolsIn{\contDisc{B^1_2}}{\spec{\last{\maph}}{\others{\veca^1}}}
= \nbSolsIn{\contDisc{B^1_2}}{\spec{\last{\maph}}{\others{\veca^4}}}=3$.
Thus applying Thm.~\ref{th_count}
yields $\nbSolsIn{\vecD(\vecB^1)}{\maph}=3\times 3$.

%
Let $\vecB^2$
be the polybox centered
in $(0,1)$ having width $2\times2^{-\delta}$.
One can apply Thm.~\ref{th_count} to obtain 
$\nbSolsIn{\vecD(\vecB^2)}{\mapg}=2\times1$
and $\nbSolsIn{\vecD(\vecB^2)}{\maph}=3\times1$.
The real parts of $\vecD(\vecB^1)$ and $\vecD(\vecB^2)$
are depicted in Fig.~\ref{fig_sols}.
\medskip

\begin{proof}[of Thm.~\ref{th_count}.]
Remark that 
$\solsIn{\vecD}{\mapf}=\{ \veca\in\vecD | \mapf(\veca)=\veczero \}$
can be defined in an inductive way as
$\solsIn{\vecD}{\mapf}=
\{ (\vecb,c)\in\vecD |
   \vecb\in\solsIn{\others{\vecD}}{\others{\mapf}}
   \text{ and } 
   c\in
   \solsIn{\last{\vecD}}{\spec{\last{\mapf}}{\vecb} } \}$.
Using \refPro{zhang}, we may write

\noindent$\nbSolsIn{\vecD}{\mapf} = \sum\limits_{(\vecb,c)\in\solsIn{\vecD}{\mapf}}
      \multipli{\vecb}{\others{\mapf}}
      \times \multipli{c}{\spec{\last{\mapf}}{\vecb}}$

\noindent$\hphantom{\nbSolsIn{\vecD}{\mapf}} =
      \sum\limits_{\vecb\in\solsIn{\others{\vecD}}{\others{\mapf}}}
        \left(
        \multipli{\vecb}{\others{\mapf}}
        \times
     \sum\limits_{c\in\solsIn{\last{\vecD}}{\spec{\last{\mapf}}{\vecb}}}
         \multipli{c}{\spec{\last{\mapf}}{\vecb}}
         \right)$
         
\noindent$\hphantom{\nbSolsIn{\vecD}{\mapf}} = 
     \sum\limits_{\vecb\in\solsIn{\others{\vecD}}{\others{\mapf}}}
        \multipli{\vecb}{\others{\mapf}}
        \times m$
        
\noindent$\hphantom{\nbSolsIn{\vecD}{\mapf}} = 
		m\times\nbSolsIn{\others{\vecD}}{\others{\mapf}}$.
		
\qed
\end{proof}

\subsection{Tower Representation}
\label{subsection_TAC}

\begin{definition}[Tower Representations]
\label{defi_TAC}
		Let $\vecD\ib\C^\dim$ be a polydisc
		and $\vecm$ a $n$-vector of positive integers.
		The pair $(\vecD,\vecm)$
		is a \dt{tower} (relative to $\mapf$) if it satisfies:
		if $\dim=1$, then $(\vecD,\vecm)=(\vecD_1,\vecm_1)$ is
		a cluster representation relative to $\mapf$.
		Inductively, if $\dim>1$ then:
 \begin{enumerate}[$(i)$]
  \item $(\others{\vecD},\others{\vecm})$ is a tower 
		  relative to $\others{\mapf}$
  \item $\forall \vecb\in\others{\vecD}$, 
		  $(\last{\vecD},\last{\vecm})$ is a cluster representation 
				 relative to
				 $\spec{\last{\mapf}}{\vecb}$.
 \end{enumerate}
	The \dt{height} of the tower $(\vecD,\vecm)$ is $\dim$.
\end{definition}

If we replace `cluster' by `natural cluster'
in the above definition, then $(\vecD,\vecm)$ a \dt{natural tower}.
If $\vecB$ is a polybox, and $(\contDisc{\vecB},\vecm)$
is a tower relative to $\mapf$, then we can also
call $(\vecB,\vecm)$ a \dt{(polybox) tower} relative to $\mapf$.
Below, we will only consider natural towers and will omit
the word natural.
\medskip

\ignore{
Note that if $(B,m,g)$ is an algebraic cluster,
$\solsIn{\contDisc{B}}{g}$ is a natural cluster.
One can easily see with an inductive reasoning that 
if $(\vecB,\vecm,\mapf)$ is a TAC, 
$\nbSolsIn{\contDisc{\vecB}}{\mapf}=
\nbSolsIn{3\contDisc{\vecB}}{\mapf}$
and $\contDisc{\vecB}$ is a natural cluster of solutions
of $\mapf$.
\medskip
}%

Let $\mapg,\maph$ be defined 
as in Eqs.~\ref{eq_syst_1} and~\ref{eq_syst_2}
and $\vecB^1=(B^1_1,B^1_2)$, $\vecB^2=(B^2_1,B^2_2)$ be
as defined in \ref{subsection_multiplicities}.
The pair $(\contDisc{B^1_1},3)$ is a tower relative to $h_1$.
Moreover, if $\delta\geq 3$, 
	$(\contDisc{\vecB^1}, (3,3))$
	and $(\contDisc{\vecB^2}, (3,1))$ are towers relative to $\maph$.
Consider the polynomial
$h_2(z_1,z_2)=(z_2+2^{\delta}z_1^2)^2(z_2-1)z_2$.
If $z_2\in 3\contDisc{B^1_2}$ then $|z_2|<\frac{3\times3}{16}<1$ and 
for any $z_1\in \contDisc{B^1_1}$, $h_2$ has 3 roots
counted with multiplicity in $\contDisc{B^1_2}$
and in $3\contDisc{B^1_2}$.
Hence for any $b\in \contDisc{B^1_1}$, 
$(\contDisc{B^1_2},3)$ is a tower relative to $\spec{h_2}{b}$
then $(\contDisc{\vecB^1}, (3,3))$
is a tower relative to $\maph$.
Similarly, $(\contDisc{\vecB^2}, (3,1))$ is a tower relative to $\maph$.
$(\vecB^1, (3,3))$ and $(\vecB^2, (3,1))$ are (polybox) towers relative to $\maph$.

In contrast, although $(\contDisc{B^1_1},2)$
is a tower relative to $g_1$,
there exist no tower relative to $\mapg$ having
$\vecB^1$ or $\vecB^2$ as box:
	$-2^{-\delta}$,
	0 and $2^{-\delta}$ are three points of 
	$B^1_1=B^2_1$; consider 
	the three polynomials 
	$\spec{g_2}{-2^{-\delta}}$, 
	$\spec{g_2}{0}$ and 
	$\spec{g_2}{2^{-\delta}}$. 
	$\spec{g_2}{-2^{-\delta}}$ and 
	$\spec{g_2}{2^{-\delta}}$ have each 
	1 root of multiplicity 1 in 
	$B^1_2$
	while 
	$\spec{g_2}{0}$ has 1 root of multiplicity $2$
	in 
	$B^1_2$:
	there is no $\vecm$ that satisfy
	condition $(ii)$ of Def.~\ref{defi_TAC}.
	In the case of $\vecB^2$,
	$\spec{g_2}{-2^{-\delta}}$ and $\spec{g_2}{2^{-\delta}}$ have both 
	1 root of multiplicity 1 in $B^2_2$
	while $\spec{g_2}{0}$ has no root in $B^2_2$.
	
\medskip

An immediate consequence of the previous theorem is

\begin{coro}
	\label{th_TAC}
    Let $(\vecD,\vecm)$ be a tower relative
	to $\mapf$ of height $n>1$. 
    Then 
    
    \[
     \nbSolsIn{\vecD}{\mapf} = \last{\vecm}\times
     \nbSolsIn{\others{\vecD}}{\others{\mapf}}.
    \]
		Inductively, we have
		$\nbSolsIn{\vecD}{\mapf} = \prod_{i=1}^n \vecm_i$
\end{coro}
\ignore{
\begin{proof}
Let $(\vecB,\vecm,\mapf)$ be a TAC
and $\veca^1,\ldots,\veca^l$ be the solutions in 
the cluster
$\solsIn{\contDisc{\vecB}}{\mapf}$.
From the definition of a TAC, one has
$\forall 1\leq i \leq l$, 
$(\last{\vecB},\last{\vecm},\spec{\last{\mapf}}{\others{\veca^i}})$ 
is an algebraic cluster
and $\nbSolsIn{\contDisc{\last{\vecB}}}{\spec{\last{\mapf}}{\others{\veca^i}}}=
\last{\vecm}$.
One can then apply Thm.~\ref{th_count} to obtain the
consequence of Thm.~\ref{th_TAC}.
\qed
\end{proof}
}

Remark finally that if $(\vecB,\vecm)$ is a tower relative to 
$\others{\mapf}$ and $f$ is an oracle for 
$\spec{\last{\mapf}}{\vecb}$ for any $\vecb\in\contDisc{\vecB}$,
one can use $Cluster$, as specified in Prop.~\ref{pro:unisolve}, 
to compute clusters of $\spec{\last{\mapf}}{\vecb}$
for any $\vecb\in\contDisc{\vecB}$ in a box $B$.
If this returns a list $\{ (B^j,m^j) | 1\leq j \leq l \}$,
then for all $1\leq j \leq l$,
$((\contDisc{\vecB},\contDisc{B^j}),(\vecm,m^j))$
is a tower relative to $\mapf$, and from corollary \ref{th_TAC},
$\nbSolsIn{(\contDisc{\vecB},\contDisc{B^j})}{\mapf}=m^j\times\prod_{k=1}^{n-1}\vecm_k$.
Moreover, $\solsIn{(\vecB,B)}{\mapf}\subseteq 
            \bigcup_{j=1}^{l} \solsIn{(\contDisc{\vecB},\contDisc{B^j})}{\mapf}
            \subseteq \solsIn{(2\vecB,2B)}{\mapf}$.
In other words, 
$\{ ((\contDisc{\vecB},\contDisc{B^j}),m^j\times\prod_{k=1}^{n-1}\vecm_k) | 1\leq j \leq l \}$
is a solution for the clustering problem in $(\vecB,B)$.

We show in Sec.~\ref{section_clustering} how to setup an oracle for
$\spec{\last{\mapf}}{\vecb}$ for any $\vecb\in\contDisc{\vecB}$.
This oracle may refine $(\vecB,\vecm)$ and split it into several clusters.
\section{Error Analysis of Approximate Specializations}
\label{section_suffCond}
	The proofs for this section is found in the Appendix.
	
	Given $f,\wtf\in \CC[\bfz]=\CC[\bfz\subx[n]]$ and $\bfb, \wtbfb\in\CC^n$,
	our basic goal is to bound the evaluation error 
		
		$$\|f(\bfb)-\wtf(\wtbfb)\|$$
	in terms of
			$\delta_f \as \|f-\wtf\|$
		and
			$\delta_\bfb \as \|\bfb-\wtbfb\|$.
	This will be done by induction on $n$.
	Our analysis aims not just
	to produce some error bound, but to express this error
	in terms that are easily understood, and which reveals
	the underlying inductive structure.
	Towards this end, we introduce the following \dt{$\beta$-bound}
	function: if $d$ is a positive integer and $b\in \CC$, 
		
		\beql{beta_txt}
		\beta(d,b)\as \sum_{i=0}^d |b|^i.
		\eeql

\noindent
	Let $\bfd = \bfd(f)$,
		i.e., $\bfd_i = \degree_{\bfz_i}(f)$ for each $i$.
	The \dt{support} of $f$ is $\supp(f)\ib\NN^n$ where
		$f = \sum_{\bfalpha\in \supp(f)} c_\bfalpha \bfz^\bfalpha$
	where $c_\bfalpha\in\CC\setminus \set{0}$. Here,
		$\bfz^\bfalpha\as \prod_{i=1}^n \bfz_i^{\bfalpha_i}$.
	We assume that $\supp(\wtf)\ib\supp(f)$.
	Our induction variable is $k=1\dd n$.
	For $\bfalpha\in\NN^n$, let
		$\pi_k(\bfalpha)\as (0\dd 0,\bfalpha_{k+1}\dd \bfalpha_n)$.
	E.g., if $k=n$ then $\pi_k(\bfalpha)=\0$.
	Thus $\bfalpha -\pi_k(\bfalpha) = 
		(\bfalpha_1\dd \bfalpha_k,0\dd 0)$.
	Next define
		$\supp_k(f)\as \set{\pi_k(\bfalpha): \bfalpha\in\supp(f)}$.
	With this notation, we can write 
		
		\beql{fk_txt}
		f = \sum_{\bfalpha\in \supp_k(f)}
			f_\bfalpha \bfz^\bfalpha
			\eeql
	where each $f_\bfalpha\in\CC[\bfz\subx[k]]$.
	E.g., if $k=n$ then $\supp_k(f)=\set{\0}$ and so $f_{\0}=f$.
	Assume that we are given $f,\wtf\in\CC[\bfz]=\CC[\bfz\subx]$
	and $\bfb,\wtbfb\in\CC$.  Also the degree sequences satisfies 
		$\bfd(\wtf)\le \bfd(f)$, that is the inequality holds
                componentwise.  Then we may define these quantities 
		for $k=1\dd n$:
		
		\beqarrys
			\delta_k \bfb &\as& 
				|\bfb_k -\wtbfb_k|, \\
			\delta_k f &\as& 
				\|f(\bfb\subx[k])-\wtf(\wtbfb\subx[k])\|
					& \qquad(\textrm{with } \delta_0 f=\|f-\wtf\|),\\
			\bfbeta_k &\as& 
				\beta(\bfd_k, \bfb_k) \\
			\wtbfbeta_k &\as& 
				\beta(\bfd_k, |\bfb_k| + \delta_k\bfb).
		\eeqarrys
		Note that $\delta_k$ is a operator that must
		attach to some function $f$ or vector $\bfb$ to
		denote the ``$k$th perturbation'' of $f$ or $\bfb$.

	\bleml{n>=1_txt}\ \\
		Let $n\ge 1$ and $k=1\dd n$:
			    \begin{center}
		\begin{tabular}{crlllll}
			(i) & $\| f(\bfb\subx[k])-f(\bfb\subx[k-1])(\wtbfb_k) \|$
				& $\le$
				& $\delta_k \bfb \cdot \|\del[k]f(\bfb\subx[k-1])\|
					\cdot \wtbfbeta_k$. \\
			(ii) & $\|f(\bfb\subx[k-1])(\wtbfb_k)-\wtf(\wtbfb\subx[k])\|$
				& $\le$
				& $\delta_{k-1} f \cdot \wtbfbeta_k$.\\
			(iii) & $\delta_k f$
				& $\le$
				& $\Big[ \delta_k \bfb \cdot \|\del[k]f(\bfb\subx[k-1])\|
					+ \delta_{k-1} f \Big]\cdot \wtbfbeta_k$.
		\end{tabular}
	    \end{center}
	\eleml

	We now have a recursive bound $\|\delta_n f\|$.
	But we need to convert the bound to only depend on the
	data $\|\bfb\|, \|f\|, \delta_k\bfb$.
	In particular, we remove any occurrences of
	$\del[k]f_\bfalpha$ with the help of the next lemma:

	\bleml{evalbound_txt}
		For $k=1\dd n$:
		\ \\(i)
		$\|f(\bfb\subx[k])\|
			\le \|f\|\cdot \prod_{i=1}^k \bfbeta_i$
		\\(ii)
		For $\bfalpha\in\supp_k(f)$,
		\begin{center}
		\begin{tabular}{crlllll}
			$\Big\|\del[k]f_\bfalpha(\bfb\subx[k-1])\Big\|$
				& $\le$
				& $\bfd_k\cdot \|f_\bfalpha(\bfb\subx[k-1])\|$.
		\end{tabular}
		\end{center}
		(iii)
		$\|\del[k]f(\bfb\subx[k-1])\| \le \bfd_k
				\cdot \|f\|\cdot \prod_{i=1}^{k-1} \bfbeta_i$
	\eleml
	
	Putting it all together:
	\bthml{deltaf_txt}
	For $k=1\dd n$,
	
	$$\delta_k f \le
			\Big[ \delta_0 f + \|f\|
				\cdot \sum_{i=1}^k \bfd_i\cdot \delta_i \bfb\Big]
				\cdot \Big(\prod_{i=1}^k \wtbfbeta_i\Big).$$
	\ethml

	The next lemma answers the question:
	given $\delta_L>0$,
	how can we ensure that 
		
		$$\delta_{n-1}f \as
			\|f(\bfb\subx[n-1])-\wtf(\wtbfb\subx[n-1])\|$$
	is upper bounded by $\delta_L$?
	\bleml{marc_txt} \ \\
		\begin{tabular}{lrllll}
			\multicolumn{3}{l}{
				Given $\delta_L>0$,
				$f,\wtf\in\CC[\bfz]$ and
				$\bfb,\wtbfb\in\C^{\dim-1}$ where $\dim>1$.}\\
			\multicolumn{3}{l}{
				Let $d=\max(\degree_{\bfz_i}(f))$ and $M=\|\bfb\|+1$.}\\
	  	If\\
			& $\delta_f $
				& $\leq \frac{\delta_L}{2((d+1)M^d)^{n-1}}$
				& (*)\\
			and\\
			& $\delta_\bfb $
				& $\leq \min(1,
					\frac{\delta_L}{2d\|f\|(n-1)((d+1)M^d)^{n-1}})$,
				& (**)\\
			then\hspace*{3mm}\\
			& $\delta_{n-1} f$
				& $\leq \delta_L.$
		\end{tabular}
	\eleml
	
\section{Clustering for Triangular Systems}
\label{section_clustering}
\input{sec4newnew}

\section{Implementation and benchmarks}
	\label{sec_benchmark}
	
	We implemented in \Julialang\footnote{\url{
	https://julialang.org/}}
	our
	complex solution clustering algorithm
	and made it available through the package
	\CCJL \footnote{\url{
	https://github.com/rimbach/Ccluster.jl}}.
	It is named hereafter \tcluster.
	It uses, as routine for clustering 
	roots of univariate polynomials
	given by approximations, 
	the univariate solver \ccluster
	described in \cite{ICMSpaper}
	and available in \CCJL.
	The procedure 
	for approximating a multivariate polynomial
	specialized in a cluster of fibers
	relies on the ball arithmetic library \texttt{arb} (see \cite{Johansson2017arb}),
    interfaced in \Julialang through the package 
    \texttt{Nemo}\footnote{\url{http://nemocas.org/links.html}}.
	
	Sec.~\ref{subsec_syst_with_clusters} reports
	how \tcluster performs on systems having clusters
	of solutions.
	Sec.~\ref{subsec_benchmark} proposes benchmarks
	for solving random dense triangular systems with only
	regular solutions, and with solutions with multiplicities;
	\tcluster is compared with three homotopy solvers.
	Sec.~\ref{subsec_regular} is about
	using \tcluster to cluster solutions
	of system triangularized with regular chains.
	Unless specified, \tcluster is used 
	with $\epsilon=2^{-53}$.
	\tcluster global (resp. local) 
	holds for \tcluster with initial box $\vecB$
	centered in $\veczero$ with width $10^6$
	(resp. 2).
	
	All the timings given below are sequential times in seconds
	on a Intel(R) Core(TM) i7-7600U CPU @ 2.80GHz machine with linux.
	
\subsection{Clustering ability}
	\label{subsec_syst_with_clusters}
		Consider the triangular systems 
		$\mapg=(f,g_2)=\veczero$ and 
		$\maph=(f,h_2)=\veczero$ 
	    where 
	    
	    \begin{equation}
	     \begin{array}{lcl}
	      f(z_1) & = & z_1^{d_1} - (2^{\delta}z_1-1)^c\\
	      g_2(z_1,z_2) & = & z_2^{d_2}z_1^{d_2} -1 \\
	      h_2(z_1,z_2) & = & z_2^{d_2} - z_1^{d_2}
	     \end{array}
	    \end{equation}
    with $d_1=30$, $c=10$, $\delta=128$ and $d_2=10$.
    All the roots of $f$ have multiplicity 1.
    A cluster $S_1$ of 10 roots is in a disk centered in $2^{-\delta}$
    with radius $2^{-b}=2^{-\frac{d_1\delta+\delta-1}{c}}\simeq2^{-397}$
    (see \cite{Mignotte1995}).
    Since $d_1>c>1$, roots in $S_1$ have
    modulus $\leq 2^{-\delta} + 2^{-b}\leq 2^{-\delta+1}=2^{-127}=\hat{\gamma}$.
    $S_2$ denotes the set of $d_1-c$ others roots of $f$,
    that have a modulus of the order
    of $\gamma=2^{\frac{c\delta}{d_1-c}}=2^{64}$.
    The $d_2$ roots of $g_2$ are 
    on a circle centered in $0$
    with radius $\geq\hat{\gamma}^{-1}$
    when $z_1\in S_1$,
    and of order $\gamma^{-1}$
    when $z_1\in S_2$.
    The $d_2$ roots of $h_2$ are
    on a circle centered in $0$
    with radius $\leq\hat{\gamma}$
    when $z_1\in S_1$,
    and of order $\gamma$
    when $z_1\in S_2$.
    All the solutions of $\mapg=\veczero$ and 
	$\maph=\veczero$
    are included in the box 
    $\vecB$ centered in $\veczero$ with width $10^{40}$.
    
    We computed clusters of solutions for the two 
    systems with \tcluster in $\vecB$
    for four values of $\epsilon$
    and reported the cluster structure as a sum
    where $c_1$ (respectively $c_2$, $c_3$) 
    stands for the number of 
    clusters with sum of multiplicities $1$ (resp. $10$, $100$).
    Table.~\ref{table_clus} gives
    this structure in columns \#Sols,
    the solving time in columns t and
    the min and max precision required
    on clusters of $f_1$ in columns M and m
    (\emph{i.e.} the $log_2$ of the radius of the disk isolating
    the clusters).
      
    $(\mapg=\veczero)$
    has $20$ clusters of 
    $10$ solutions above each root in $S_2$,
    where solutions have pairwise distance $\simeq2^{-64}$.
    It has $10$ clusters of $10$ solutions
    above the cluster $S_1$
    where solutions have pairwise distance $\leq2^{-b}\simeq2^{-397}$.
    This structure is found by \tcluster with $\epsilon=2^{-53}$.
    When $\epsilon=2^{-106}$, the $20$ clusters above
    roots in $S_2$ are split, not the ones above 
    roots in $S_1$.
    When $\epsilon=2^{-212}$, the clusters 
    above roots in $S_1$ are split even if the
    pairwise distances between solutions in these clusters
    are far smaller than $2^{-212}$;
    this is because isolating roots 
    of $g_2$ with error less than $\epsilon=2^{-212}$
    requires more precision on roots of $f$,
    as shown is column (m,M).
    When $\epsilon=2^{-424}$,
    all the clusters are split.
    
    $(\maph=\veczero)$
    has $200$ solutions
    above roots in $S_2$
    and a cluster of $100$ simple solutions 
    above roots in $S_1$.
    The first (resp. second) components of the solutions in this cluster
    are in a disc of radius $\leq2^{-b}\simeq2^{-397}$
    (resp. $\hat{\gamma}=2^{-127}$).
    This cluster structure is found by \tcluster
    with $\epsilon=2^{-53}$ and $\epsilon=2^{-106}$.
    When $\epsilon=2^{-212}$,
    the cluster of $100$ solutions
    is split in $10$ clusters of $10$
    solutions.
    When $\epsilon=2^{-424}$, the cluster
    is split in $100$ solutions.
    
\begin{table}[t]
    \begin{center}
	\begin{scriptsize}
	\begin{tabular}{l||c|c|c||c|c|c||}
		&\multicolumn{3}{c||}{$\mapg=\veczero$} 
			&\multicolumn{3}{c||}{$\maph=\veczero$}       \\\cline{2-7}
		$log_2(\epsilon)$&  \#Sols & t (s) & (m,M)
			& \#Sols  & t (s) & (m,M) \\\hline\hline
		  -53 & $~~0 + 30\times 10$ & 0.17 & ( -212,- 424) & $200 + ~0\times 10 + 1\times 100$ & 0.54 & ( -212, -212) \\ 
		 -106 & $200 + 10\times 10$ & 0.64 & ( -212,- 424) & $200 + ~0\times 10 + 1\times 100$ & 0.57 & ( -212, -424) \\ 
		 -212 & $300 + ~0\times 10$ & 3.91 & ( -424,- 848) & $200 + 10\times 10 + 0\times 100$ & 0.66 & ( -212, -848) \\ 
		 -424 & $300 + ~0\times 10$ & 3.87 & ( -848,-1696) & $300 + ~0\times 10 + 0\times 100$ & 3.78 & ( -848, -848) \\
\end{tabular}
	\end{scriptsize}
	\caption{
	Clustering the solutions of systems defined in 
	Sec.~\ref{subsec_syst_with_clusters}
	with $d_1=30$, $c=10$, $\delta=128$, $d_2=10$
	for four values of $\epsilon$ 
	in box 
	$\vecB$ centered in $\veczero$ with width $10^{40}$.
	         }
	\label{table_clus}
	\end{center}
	\end{table}
	
	\subsection{Benchmarks with random dense systems}
	\label{subsec_benchmark}
	
	We present benchmarks 
	for randomly generated triangular systems 
	without and with
	multiple solutions.
	We compare the efficiency and the robustness of \tcluster and two
	homotopy solvers.
	
	\paragraph{Homotopy solvers.}
	Homotopy solving is a two-step process.
	First, an upper bound $D$ (either the B\'ezout's bound,
	or a bound obtained with \emph{polyhedral homotopy}, see 
	\cite{huber1995polyhedral})
	on the number of solutions of the system  is computed.
	Then $D$ paths are followed to find the solutions.	
	\highlightModifs{
	Among available homotopy solvers\footnote{other 
	major homotopy solvers are 
	{\tt NAG4M2} (for {\tt Macaulay2}), {\tt PHCpack}
	and {\tt HOM4PS-3}.
	{\tt Bertini2} is still in development.
	}, we used in our benchmarks 
	\homfps \footnote{
	\url{http://www.math.nsysu.edu.tw/~leetsung/works/HOM4PS_soft.htm}},
	\bertini\footnote{\url{https://bertini.nd.edu/}} 
	(see \cite{BHSW06})
	and \HCfnJL\footnote{
	\url{https://www.juliahomotopycontinuation.org/}}
	(hereafter, we denote it \HCJL).
	\homfps and \HCJL implement polyhedral homotopy, 
	thus follow possibly less paths.
	\bertini and \HCJL compute the multiplicity structure of solutions.
	\bertini can use an Adaptive Multi-Precision (AMP) arithmetic;
	below \bertiniAMP  refers to \bertini
	with AMP.
	}
	
	\paragraph{Systems.}
    We follow the approach of \cite{cheng2009complete}
    to generate triangular systems with and without multiple solutions.
	The \emph{type} of a 
	triangular 
	system 
	$\mapf(\vecz)=\veczero$ with $\dim$ equations 
	is the list $(\deg_1,\ldots,\deg_\dim)$
	where $\deg_i=deg_{z_i}(\mapf_i)$.
	A random dense polynomial $\mapf_i\in\C[\vecunkdim{z}{i}]$ 
	of degree $\deg_i$ in $z_i$ 
	is generated as follows.
    If $i>1$, $\mapf_i= \sum_{j=0}^{\deg_i} g_j z_i^j$
	where $g_j\in\C[\vecunkdim{z}{i-1}]$
	is a random dense polynomial of degree $\deg_i-j$ in $z_{i-1}$. 
    $\mapf_1$ is a random dense polynomial in $\C[z_1]$ of degree $\deg_1$.
	A system $\mapf(\vecz)=\veczero$ 
	of type $(\deg_1,\ldots,\deg_\dim)$
	is obtained by generating successively 
	random dense polynomials $\mapf_i$ of degrees $\deg_i$ in $z_i$.
	Triangular systems with multiple solutions 
	are obtained by taking 
	$\mapf_1$ as above, and for $i=2,\ldots,\dim$,
	$\mapf_i=
	a_i^2(b_iz_i+c_i)^{\lfloor\frac{\deg_i+1}{2}\rfloor-\lfloor\frac{\deg_i}{2}\rfloor}$
	where $a_i\in\C[\vecunkdim{z}{i}]$ has degree $\lfloor\frac{\deg_i}{2}\rfloor$ in $z_i$ and 
	$b_i,c_i$ are in $\C[\vecunkdim{z}{i-1}]$ and have degrees
	$\deg_i$ in $z_{i-1}$.
	
	\paragraph{Benchmarks.}
	In Table~\ref{table_tri_random},
	we compare 
	the three homotopy solvers
	and \tcluster 
	global and local
	on triangular systems
	with integer coefficients 
	without and with multiple solutions.
	Coefficients of systems without multiple solutions are in $[-2^{9},2^{9}]$,
	while coefficients of systems with multiple solutions are in $[-2^{34},2^{34}]$.
	In both cases, we generated $5$ systems of each type.
    Here \tcluster global 
    found all the solutions
	but in general this is not guaranteed.
	The columns \#Sols give the average number of solutions counted with
        multiplicities found 
	by each solver and the columns t
	the average time.
	The columns \#Clus give the average number of clusters found by \tcluster.
	The systems we generated have $\deg_1\times\ldots\times \deg_\dim$ 
	solutions which is the B\'ezout's bound, 
	and the homotopy solvers have to follow this number of paths.
	
	\paragraph{Systems with only simple solutions.} 
	For type (9,9,9,9,9), \bertiniAMP has been stopped after 1 hour and
        \homfps terminates with a segmentation fault.
	Homotopy solvers should find all the solutions.
	\bertiniAMP failed in this task for one system of type
	$(9,9,9,9)$ 
	\highlightModifs{and two systems of type
	$(2,2,2,2,2,2,2,2,2,2)$}
	but acknowledged that solutions could be missing.
	\homfps returns incorrect results without warnings.
	In contrast, \tcluster global always 
	finds the correct number of solutions.
	\tcluster global is in general faster than \bertiniAMP
	and is faster than \homfps for systems
	of types $(6,6,6,6,6)$ and $(9,9,9,9)$.
	\highlightModifs{For systems of highest degree polynomials, 
	\tcluster global and \HCJL present similar solving times.
	The timings for systems of type $(2,2,2,2,2,2,2,2,2,2)$
	emphasize that the efficiency of our solver is not
	penalized by high dimensional systems since it performs
	inductively subdivisions in boxes in $\CC$.}
	\tcluster local is significantly faster than the other
	approaches.

\begin{table}[t]
	\begin{center}
	\begin{scriptsize}
	\begin{tabular}{c||c|c||c|c||c|c||c|c||c|c||}
            &\multicolumn{2}{c||}{\tcluster local}
            &\multicolumn{2}{c||}{\tcluster global}
            &\multicolumn{2}{c||}{\homfps}
            &\multicolumn{2}{c||}{\bertiniAMP}
            &\multicolumn{2}{c||}{\HCJL}\\\hline
	type        &  \#Sols, \#Clus  & t (s) & \#Sols, \#Clus & t (s) & \#Sols & t (s) & \#Sols & t (s)& \#Sols & t (s)\\\hline
	\multicolumn{9}{l}{Systems with only simple solutions}\\\hline
(6,6,6)     &  34.2, 34.2  & 0.04  &  216,   216   & 0.35          & 216          & \coblue{0.06}  & 216           & 1.17    & 216   & 2.77  \\\hline
(9,9,9)     &  149,  149   & 0.24  &  729,   729   & 1.43          & \cored{713}  & \coblue{0.47}  & 729           & 29.3    & 729   & 4.21  \\\hline
(6,6,6,6)   &  63.4, 63.4  & 0.10  &  1296,  1296  & 2.21          & \cored{1274} & \coblue{1.37}  & 1296          & 24.2    & 1296  & 4.70  \\\hline
(9,9,9,9)   &  559, 559    & 1.06  &  6561,  6561  & 14.6          & \cored{6036} & 111            & \cored{6560}  & 1605    & 6561  & \coblue{14.0}  \\\hline
(6,6,6,6,6) &  155, 155    & 0.37  &  7776,  7776  & 13.8          & \cored{7730} & 28.6           & 7776          & 318     & 7776  & \coblue{11.5}  \\\hline
(9,9,9,9,9) &  1739, 1739  & 4.83  &  59049, 59049 & 130           & -            & -              & ?             & $>$3600 & 59049 & \coblue{116}   \\\hline
(2,2,2,2,2,2,2,2,2,2)& 0, 0 & 0.13 & 1024, 1024 & 2.92 & 1024 & \coblue{2.74} & \cored{1023} & 8.63 & 1024 & 4.84\\\hline
	\multicolumn{9}{l}{Systems with multiple solutions}\\\hline
	(6,6)  &   10.8, 5.40    &  0.01 &  36, 18     & 0.06          &  36            & \coblue{0.00}  &  18    & 3.63    & \cored{17.4} & 1.74  \\\hline
	(9,9)  &   23.8, 13.6    &  0.03 &  81, 45     & 0.17          &  \cored{67.4}  & \coblue{0.06}  &  45    & 218     & \cored{33.6} & 3.27    \\\hline
	(6,6,6)&   35.2, 8.80    &  0.05 &  216, 54    & 0.26          &  \cored{210}   & \coblue{0.16}  &  54    & 47.9    & \cored{53.2}          & 2.75    \\\hline
	(9,9,9)&   113,  37.6    &  0.22 &  729, 225   & \coblue{1.10} &  \cored{357}   & 18.9           &  ?     & $>$3600 & \cored{159}  & 28.4  \\\hline
	(6,6,6,6)&  81.6, 10.2 & 0.21 & 1296, 162 & \coblue{1.29} & \cored{1010} & 4.46 & 162 & 662 & \cored{134} & 8.06\\\hline
	\end{tabular}
	\end{scriptsize}
	\end{center}
	\caption{Solving random dense triangular systems with \tcluster, \homfps, \bertiniAMP
	        and \HCJL.
	}
	\label{table_tri_random}
\end{table}	
		
		\paragraph{Systems with multiple solutions.}
		A well isolated multiple solution
		is reported by \tcluster in a cluster with its multiplicity.
		In all cases, the number of clusters found 
		by \tcluster global is the number of distinct solutions of each
		systems.
		\homfps fails in finding all the solutions.
		\bertiniAMP computes correctly the multiplicity of solutions.
		\HCJL fails in computing correctly the multiplicity structure 
		of solutions.
		For type (9,9,9), \bertiniAMP has been stopped after 1 hour.
		\tcluster global is faster than \bertiniAMP~
		and \HCJL,
		and faster than \homfps for systems of type
		$(9,9,9)$ and $(6,6,6,6)$.
		
		\subsection{Systems obtained by triangularization}
		\label{subsec_regular}
		
		In this subsection, we report on using \tcluster
		for clustering the solutions of triangular systems $\mapf(\vecz)=\veczero$
		obtained from a non-triangular system $\mapg(\vecz)=\veczero$
		with Regular Chains (RC, see \cite{AUBRY1999105,Dahan:2005:LTT:1073884.1073901}).
		Algorithms for triangularizing systems with RC
		produce a set of triangular systems 
		$\{ \mapf_1(\vecz)=\veczero, \ldots, \mapf_l(\vecz)=\veczero \}$
		having distinct solutions
		whose union 
		is the set of distinct solutions of $\mapf(\vecz)=\veczero$.
		The multiplicities of solutions are not preserved
		by this process.
		
		\paragraph{Systems.}
		We consider non-triangular systems $\mapg(\vecz)=\veczero$ 
		both classical (coming from \cite{boulier2014}),
		and sparse random where $\mapg=(g_1,\ldots,g_\dim)$ and each $g_i$ has the form
		$g_i(\vecz) = z_i^{\deg_i} - g_i'(\vecz)$
		where $g_i'$ is a polynomial in $\Z[\vecz]$ having total degree $\deg_i-1$,
		integers coefficients in $[-2^{8},2^{8}]$ and 5 monomials.
		The \emph{type} of such a system is the tuple $(\deg_1,\ldots, \deg_\dim)$.
		The set of all the examples can be found at 
		\url{https://cims.nyu.edu/~imbach/IPY19/IPY19.txt}.

	\paragraph{The benchmark.}
	For several types,
	we generated a system as described above and computed 
	a triangular systems with the \maple function
	\texttt{RegularChains[Trian\-gularize]}
	with option \texttt{'probability'=0.9}.
	For the classical systems, we used no option.
	In table~\ref{table_nontriangular}, column RC
	gives the time to compute the RCs.
	We solved the triangular systems of the obtained 
	regular chains with \tcluster; 
	columns \tcluster global 
	report the number of solutions
	and solving time for \tcluster.
	We also used the function \texttt{RootFinding[Isolate]} of \maple
	with options \texttt{digits=15, output=interval, method='RC'}
	(\emph{i.e.} using regular chains)
	to solve our systems;
	columns Isolate RC report the number of real solutions
	and the solving time for Isolate.
	We also used \bertiniAMP to solve the original systems;
	columns \bertiniAMP report
	the number of paths followed (column \#Paths),
	the solving time and
	the number of solutions with the multiplicity structure
	found by \bertini:
	$c_1 + c_2\times m_2 + c_3\times m_3$ means
    $c_1$ (respectively $c_2$, $c_3$) 
    solutions with multiplicity 1 (resp. $m_2$, $m_3$).
	\highlightModifs{
	We also tested \homfps and \HCJL for these systems.
	The running time of \homfps is always less than 0.05s, 
	    but the number of solutions reported is wrong.
    \HCJL always finds the correct number of solutions
    but is slower than \bertiniAMP except for two
    systems 
    for which polyhedral
    homotopy allows to reduce the number of
    paths to be followed.
    \HCJL solves
    {\it Czapor-Geddes-Wang} in 3.67 s
    and {\it 5-body-homog} 
    in 3.44 s.
	}
	
\begin{table}[t]
	\begin{center}
	\begin{scriptsize}	
	\begin{tabular}{c||c|c|c||c|c||c||c|c||}
	           &
	           \multicolumn{3}{c||}{\bertiniAMP}&
	           \multicolumn{2}{c||}{Isolate RC}&
	           \multicolumn{1}{c||}{ RC}&
	           \multicolumn{2}{c||}{\tcluster global}\\\hline
	type/name  & \#Sols & \#Paths & t (s)
	           & \#Sols & t (s)
	           & t (s)
	           & \#Sols & t (s)\\\hline
	\multicolumn{9}{l|}{Random systems}\\\hline
	(4,4,4)   & 64  & 64  & \coblue{0.06} & 6 & 7.53    & 3.82 & 64  & 0.80 \\
	(5,5,5)   & 125 & 125 & \coblue{0.30} & ? & $>$1000 & 24.2 & 125 & 6.89 \\
	(3,3,3,4) & 108 & 108 & \coblue{0.13} & ? & $>$1000 & 52.4 & 108 & 3.42 \\
	(3,3,4,4) & 144 & 144 & \coblue{0.26} & ? & $>$1000 & 68.7 & 144 & 8.59 \\

	\hline
	\multicolumn{9}{l|}{Classical systems with only simple solutions}\\\hline
	
	{\it Arnborg-Lazard}     & 20 & 120 & 0.80          & 8  & 3.09 & \coblue{0.08} & 20 & \coblue{0.07}  \\
	{\it Czapor-Geddes-Wang} & 24 & 720 & 28.6          & 2  & 1.87 & \coblue{0.17} & 24 & \coblue{0.38}  \\
	{\it cyclic-5}           & 70 & 120 & \coblue{0.35} & 10 & 1.92 & 0.55          & 70 &  0.71          \\

	\hline
	\multicolumn{9}{l|}{Classical systems with multiple solutions}\\\hline
	
	{\it 5-body-homog}   & $45 + 2\times3 + 2\times24$ & 224 & 7.63          & 11 & 8.30 & \coblue{0.16} & 49  & \coblue{0.38} \\
	{\it Caprasse}       & $24 + 8\times 4 $           & 144 & \coblue{0.25} & 18 & 1.49 & 0.24          & 32  & 0.12          \\
	{\it neural-network} & $90 + 18\times 2 $          & 162 & \coblue{0.36} & 22 & 5.82 & 0.13          & 108 & 0.56          \\
	
	\end{tabular}
		\end{scriptsize}
		\end{center}
		\caption{Solving non-triangular systems with regular chains and \tcluster,
		and \bertiniAMP. 
    }
	\label{table_nontriangular}
\end{table}

\paragraph{Random systems in Table~\ref{table_nontriangular}.} 
	Here the number of solutions is the B\'ezout's
	bound and \bertiniAMP follows one path per solution.
	Homotopy solving in these cases is much more efficient 
	than triangularizing the system with RC. 
	The RC algorithm produces
	a triangular system
	of type $(d,1,\ldots,1)$ where $d$ is the B\'ezout's
	bound
	with a huge bitsize:
	For the type $(3,3,4,4)$,
	the triangular system has type $(144,1,1,1)$ 
	and each equation has bitsize about $738$.
	\tcluster has to isolate some solutions
	of the first equation at precision $2^{-424}$.
	Solving the first equation with 
	\ccluster and $\epsilon=2^{-424}$ takes 
	$8.27$s:
	\tcluster spends most of the time in
	isolating roots of the first polynomial.
	Any improvement of \ccluster will directly benefit to \tcluster.
	For three of these systems, \texttt{RootFinding[Isolate]}
	has been stopped after 1000s.
	
	\paragraph{Classical systems with only simple solutions in Table~\ref{table_nontriangular}.}
	These systems have few finite solutions compared to their B\'ezout's
        bounds, and \bertiniAMP wastes time in following paths going to infinity.
	In contrast, \tcluster is sensitive to the number of solutions
	in the initial solving domain.
	This explains why computing triangular systems and
	solving it with \tcluster is faster than \bertiniAMP for systems
	{\it Arnborg-Lazard}, {\it Czapor-Geddes-Wang}.

	\paragraph{Classical systems with multiple solutions in Table~\ref{table_nontriangular}.}
	For these systems, \bertiniAMP reports the multiplicity structure
	of the solutions.
	The triangularization step removes the multiplicity, and the RCs
	obtained are easier to solve; 
	\tcluster finds only clusters with one solution counted with
	multiplicity.

\highlightModifs{
\section{Future work}
We presented an algorithm for computing clusters of complex solutions, together with multiplicity information, 
of triangular systems of polynomial equations.
It is numerical and certified, it handles solutions with multiplicity 
and works locally. It can deal with systems whose equations are given 
by oracle polynomials.
An implementation is publicly available and the experiments we carried
out show the efficiency and robustness of our approach.

Our error analysis for the partial specialization of polynomials on 
algebraic numbers represented by oracle numbers is a first 
step towards a complexity analysis of our algorithm, which 
constitutes our future work.
We would like to present such an analysis in terms of geometric parameters
(\emph{e.g.} separation of solutions) instead of only syntactic parameters
(bit-size and degree).
}

\bibliographystyle{splncs04}
\bibliography{references}
\newpage
\section*{Appendix : Error Analysis}
	\input{app1}


\end{document}

%% file: sols2.pdf_t
\begin{picture}(0,0)%
\includegraphics{sols2.pdf}
\end{picture}%
\setlength{\unitlength}{4144sp}%
\begingroup\makeatletter\ifx\SetFigFont\undefined%
\gdef\SetFigFont#1#2#3#4#5{%
  \reset@font\fontsize{#1}{#2pt}%
  \fontfamily{#3}\fontseries{#4}\fontshape{#5}%
  \selectfont}%
\fi\endgroup%
\begin{picture}(4595,2760)(-827,59)
\put(-89,1874){\makebox(0,0)[lb]{\smash{{\SetFigFont{10}{12.0}{\rmdefault}{\mddefault}{\updefault}{\color[rgb]{0,0,0}$1$}%
}}}}
\put(2656,1874){\makebox(0,0)[lb]{\smash{{\SetFigFont{10}{12.0}{\rmdefault}{\mddefault}{\updefault}{\color[rgb]{0,0,0}$1$}%
}}}}
\put(1846,839){\makebox(0,0)[rb]{\smash{{\SetFigFont{10}{12.0}{\rmdefault}{\mddefault}{\updefault}{\color[rgb]{0,0,0}$2^{-\delta}$}%
}}}}
\put(2206,1109){\makebox(0,0)[rb]{\smash{{\SetFigFont{10}{12.0}{\rmdefault}{\mddefault}{\updefault}{\color[rgb]{0,0,0}$\textbf{a}^4$}%
}}}}
\put(-539,1109){\makebox(0,0)[rb]{\smash{{\SetFigFont{10}{12.0}{\rmdefault}{\mddefault}{\updefault}{\color[rgb]{0,0,0}$\textbf{a}^4$}%
}}}}
\put(271,884){\makebox(0,0)[lb]{\smash{{\SetFigFont{10}{12.0}{\rmdefault}{\mddefault}{\updefault}{\color[rgb]{0,0,0}$\textbf{a}^1$}%
}}}}
\put(2926,884){\makebox(0,0)[lb]{\smash{{\SetFigFont{10}{12.0}{\rmdefault}{\mddefault}{\updefault}{\color[rgb]{0,0,0}$\textbf{a}^1$}%
}}}}
\put(2206,1874){\makebox(0,0)[rb]{\smash{{\SetFigFont{10}{12.0}{\rmdefault}{\mddefault}{\updefault}{\color[rgb]{0,0,0}$\textbf{a}^3$}%
}}}}
\put(-494,1874){\makebox(0,0)[rb]{\smash{{\SetFigFont{10}{12.0}{\rmdefault}{\mddefault}{\updefault}{\color[rgb]{0,0,0}$\textbf{a}^3$}%
}}}}
\put(2971,1649){\makebox(0,0)[lb]{\smash{{\SetFigFont{10}{12.0}{\rmdefault}{\mddefault}{\updefault}{\color[rgb]{0,0,0}$\textbf{a}^2$}%
}}}}
\put(271,1649){\makebox(0,0)[lb]{\smash{{\SetFigFont{10}{12.0}{\rmdefault}{\mddefault}{\updefault}{\color[rgb]{0,0,0}$\textbf{a}^2$}%
}}}}
\put(3016,524){\makebox(0,0)[lb]{\smash{{\SetFigFont{10}{12.0}{\rmdefault}{\mddefault}{\updefault}{\color[rgb]{0,0,0}$\textbf{a}^6$}%
}}}}
\put(2161,524){\makebox(0,0)[rb]{\smash{{\SetFigFont{10}{12.0}{\rmdefault}{\mddefault}{\updefault}{\color[rgb]{0,0,0}$\textbf{a}^5$}%
}}}}
\put( 46,389){\makebox(0,0)[lb]{\smash{{\SetFigFont{10}{12.0}{\rmdefault}{\mddefault}{\updefault}{\color[rgb]{0,0,0}$2^{-\delta}$}%
}}}}
\put(-269,389){\makebox(0,0)[rb]{\smash{{\SetFigFont{10}{12.0}{\rmdefault}{\mddefault}{\updefault}{\color[rgb]{0,0,0}$2^{-\delta}$}%
}}}}
\put(2386,299){\makebox(0,0)[rb]{\smash{{\SetFigFont{10}{12.0}{\rmdefault}{\mddefault}{\updefault}{\color[rgb]{0,0,0}$2^{-\delta}$}%
}}}}
\put(2791,299){\makebox(0,0)[lb]{\smash{{\SetFigFont{10}{12.0}{\rmdefault}{\mddefault}{\updefault}{\color[rgb]{0,0,0}$2^{-\delta}$}%
}}}}
\put(2611,119){\makebox(0,0)[lb]{\smash{{\SetFigFont{10}{12.0}{\rmdefault}{\mddefault}{\updefault}{\color[rgb]{0,0,0}$-1$}%
}}}}
\put(1261,2684){\makebox(0,0)[b]{\smash{{\SetFigFont{10}{12.0}{\rmdefault}{\mddefault}{\updefault}{\color[rgb]{0,0,0}${\bf \Delta}({\bf B}^2)$}%
}}}}
\put(1171,119){\makebox(0,0)[b]{\smash{{\SetFigFont{10}{12.0}{\rmdefault}{\mddefault}{\updefault}{\color[rgb]{0,0,0}${\bf \Delta}({\bf B}^1)$}%
}}}}
\end{picture}%

%% file: sec4newnew.tex
%

    We now present our algorithm for solving the LCP
    for a given triple $(\mapf,\vecB^0,\epsilon)$,
    where $\epsilon>0$.
    Instead of $\epsilon$, we use $L\as \ceil{\log_2(1/\epsilon)}$.
    $\vecB^0=\vecB^0\subx$ is the ROI (a polybox).
    $\mapf=\mapf\subx$ is a triangular map
    with $0$-dimensional set of zeros.
    We give our algorithm in the case where
    each $\mapf_i$ is known exactly;
    it can be generalized for oracle polynomials. 
    It is based on the one-dimensional clustering algorithm
    (see Prop.~\refPro{clussolve}),
    that proceeds by subdividing the initial ROI.
    The key 
    stone in such a subdivision algorithm
    is a test that counts the number of roots with multiplicity
    in a disk. In the one-dimensional case, it is
    done with the so-called Pellet's test.
    Here we use this test with interval polynomials
    to compute towers.
    The main objects manipulated in our algorithms are cluster oracles,
    and their generalization when $\dim>1$.
    
    \paragraph{Cluster oracles in dimension $\dim\geq1$.}
    
    A polybox $\vecB\ib\CC^n$ is called an \dt{$\ell$-tower}
	if there exists an $\ell$-vector $\vecm\subx[\ell]$ such that
	$(\vecB\subx[\ell],\vecm\subx[\ell])$ is a tower.
    A cluster oracle $\calO$ in dimension $\dim>1$ 
    is defined to be a triple
		
		$$\calO=\bang{\ell,\vecB,\vecL}
			=\bang{level(\calO), domain(\calO), precision(\calO)}$$
    where
	$\ell \in \{0\dd n\}$ is called the \dt{level},
	$\vecB$ is a polybox
	called the \dt{domain}
	and $\vecL$ is a vector of integers called the \dt{precision}.
    We will guarantee that if $level(\calO)\geq 1$,
    $domain(\calO)$ is an $\ell$-tower
    and $\radius{\contDisc{domain(\calO)_i}}\leq 2^{-\vecL_i}$.
    The multiplicity information
	is implicitly carried out by a cluster oracle.
	
	\paragraph{Cluster oracles at level 1.}
	
	We generalize the $ClusterOracle$ algorithm
	in \refPro{clussolve} to $ClusterOracle1(\mapf,\calO)$,
	which returns a set $\set{\calO^1\dd \calO^k}$ of
	cluster oracles at level $1$.
	If $L=precision(\calO)_1$
	and $\vecB^i$ is the domain of $\calO^i$, 
	then $(\contDisc{\vecB^i})_1$ has radius at most $2^{-L}$,
	and $\vecB^i_j=(domain(\calO))_j$ for $j=2\dd n$.
	Moreover, the
	domains of these $\calO^i$'s form a cover for the 
	solution set of $\mapf$ in the
	domain of
	$\calO$.  All our oracles are subsequently descended from
	these $\calO^i$'s.
	
    \paragraph{Pellet test.}
    
    Our goal is to ``lift'' a
    cluster oracle $\calO=\bang{\ell,\vecB,\vecL}$
    to one or more at level $\ell+1$ (provided $\ell<\dim$)
    arising from subdividing $\vecB$.
    The fundamental tool for this purpose is the ``Pellet
	test'' and its variants (Graeffe-accelerated,
	soft-version, etc.~ -- see
	\cite{2016Becker,BECKER2017,ICMSpaper}).
	Without distinguishing among these variants,
	we may describe a \dt{generic Pellet test} denoted
		
		$$T_*(\mapf_{\ell+1},\vecB\subx[\ell],\vecB_{\ell+1})$$
	which returns an integer $m\ge -2$. $m\ge 0$ holds only if
	$m=\nbSolsIn{\contDisc{\vecB_{\ell+1}}}{\mapf_{\ell+1}(\vecB\subx[\ell])}$,
	where $\mapf_{\ell+1}(\vecB\subx[\ell])$
	is the univariate interval polynomial obtained by evaluating 
	$\mapf_{\ell+1}$ on $\vecB\subx[\ell]$.
	If $m\ge 1$, then this implies that
	$\vecB$ is an $(\ell+1)$-tower.   
	If $m=0$
	$\vecB$ does not contain zeros of $\mapf$.
	If $m=-1$ or $m=-2$, we say that the $T_*$ test \dt{failed}.
	These two modes of failure are important to understand for
	efficiency.  Informally\footnote{
		The two failure modes may be traced to our soft comparison
		of real numbers $x:y$
		(see \cite{yap-sagraloff-sharma:cluster:13,ICMSpaper}).
		It is reduced to the interval comparison
		$(x)_L:(y)_L$ for increasing $L$.  If we
		can conclude $x>y$ or $x<y$, it is a success,
		else it is a failure.  There are two failure modes:
		if we can conclude $\half x<y< 2x$, this is a $(-1)$-failure
		(it is a potential "zero problem").
		Otherwise it is a $(-2)$-failure
		(we repeat the interval test with larger $L$).
	}
	$m=-1$ means the disc $\contDisc{\vecB_{\ell+1}}$ 
	is not well-isolated (there are zeros near its boundary).
	In this case, the response is to
	subdivide 
	$\vecB_{l+1}$.
	On the other hand, $m=-2$ means we need more
	accuracy in the evaluation 
	$\mapf_{\ell+1}(\vecB\subx[\ell])$,
	which requires subdividing components $\vecB_i$ 
	of $\vecB$ with $i<\ell$.
	
	\paragraph{Lift of a natural cluster.} 
	The lifting process is performed by a function
	\[ClusterOracleN(\mapf,\calO)\]
	that takes in input a triangular map and a cluster oracle
	and outputs a pair $(flag, S)$
	where $flag\in\{{\bf success}, {\bf failure}\}$ and
	$S$ is a set of cluster oracles.
	It is essentially the clustering algorithm depicted in 
	\cite{2016Becker};
	it uses 
	the $T_*$-test
	described above with $\ell=level(\calO)$
	to count the number of roots in a disc.
	It returns the pair $({\bf failure}, \calO)$
	when one $T_*$-test returns $-2$.
	When $ClusterOracleN(\mapf,\calO)$ 
	returns $({\bf success}, S)$,
	then $S$ is a list of pairwise disjoint 
	cluster oracles at level
	$\ell+1$ so that any solution in $\calO$
	is in a cluster oracle in $S$.

    \paragraph{Solving the LCP problem.}
	We are ready to present our main algorithm,
	called $ClusterTri(\mapf,\vecB,L)$, described in Algo.~\ref{algo:ClusterTri}.
	It uses a queue $Q$ to hold the active cluster oracles
	and lift these clusters level by level to level $\dim$.
	Let $\calO$ be a cluster oracle in $Q$.
	If $level(\calO)=0$, $\calO$ is lifted
	with $ClusterOracle1$ which 
	returns a set $S$ of cluster oracles at level 1
	containing all the solutions in $\calO$.
	If $level(\calO)>0$, $\calO$ is lifted with 
	$ClusterOracleN$, which may fail;
	in that case, the asked precision for 
	levels less that $\ell$ of $\calO$
	is doubled and it's level is set to $0$,
	this will force its refining in later executions of the 
	{\bf while} loop.
	When the lift of $\calO$ succeeds, 
	one obtains a set $S$ of cluster oracles at level
	$\ell+1$.
    
	\begin{algorithm}[t]
		\begin{algorithmic}[1]
		\caption{$ClusterTri(\mapf,\vecB,L)$}
		\label{algo:ClusterTri}
		\Require{A triangular map $\mapf=\mapf\subx$, 
		         a ROI $\vecB^0=\vecB^0\subx$ 
		         and a precision $L>1$.} 
		\Ensure{A set of cluster oracles at level $n$, that solves the LCP
		        for $(\mapf,\vecB^0,2^{-L})$.}
		\State $Q.push(\bang{0,\vecB^0,(L,\ldots, L)})$
		       \hspace{1cm}\textit{\coblue{//initial cluster oracle at level $0$}}
        \While{$Q$ contains cluster oracles at level less than $\mdim$}
            \State $\calO=\bang{\ell,\vecB,\vecL}\ass Q.pop()$ \hspace{1cm}\textit{\coblue{//assume $\ell<\mdim$}}
            \If{$\ell=0$}
                \State $Q.push(ClusterOracle1(\mapf,\calO))$
            \Else
                \State $\{flag, S\}\ass ClusterOracleN(\mapf,\calO)$
                \If{$flag={\bf success}$}
                    \State $Q.push(S)$ \hspace{1cm}\textit{\coblue{//$S$ is a set of cluster oracles at level $\ell+1$}}
                \Else
                    \State $precision(\calO)\ass (2\vecL\subx[\ell],\vecL_{\ell+1},\ldots,\vecL_\mdim)$
                    \State $level(\calO)\ass 0$
                    \State $Q.push(\calO)$ \hspace{1cm}\textit{\coblue{//$\calO$ will be refined later}}
                \EndIf
            \EndIf
		\EndWhile
		\State \Return $Q$
		\end{algorithmic}
		\end{algorithm}
	
	The correctness of $ClusterTri$ is a direct consequence of
	the correctness of $ClusterOracle1$ (see Prop.~\ref{prop_clussolve}) 
	and $ClusterOracleN$,
	and corollary \ref{th_TAC}.
	
	The halting of $ClusterTri$ is a consequence of
	\refLem{marc_txt} (equation (**)) which shows that as long as
	the radius of $\contDisc{\vecB\subx[\ell]}$ approaches zero,
	Pellet test will eventually succeed;
	thus so does $ClusterOracleN$.

\ignore{
{\bf Open Problems.}
	\\1. We have structured our description through
	cluster oracles to capture the flow of information and
	control.  The optimal management of these oracle operations
	calls for further investigation.  
	For instance, the feedback from failure of the $T_*$-test
	is currently two-valued.  It is possible to give more quantitative
	feedback in terms of additional accuracy needed.
	\\2. The set oracles can be organized into a
	tree (if $\calO$ at level $k-1$
	and $\calO'$ at level $k$ have domains $\bfDelta$
	and $\bfDelta'$ such that $\bfDelta\subx[k-1]=\bfDelta'\subx[k-1]$,
	than $\calO'$ is the child of $\calO$).
	\\3. Generalize the Newton search method to $\bfDelta\subx[k-1]$.
}%


%% file: app1.tex
\label{subsection_suffCond}
\renewcommand{\moveToApp}[1]{#1}

	\progb{\lline
	\lline
	\lline[2] This Appendix contains all the proofs for
	our error analysis.
	\lline[2] Section 3 is an excerpt.
	}

	Given $f,\wtf\in \CC[\bfz]$ and $\bfb, \wtbfb\in\CC^n$,
	our basic goal is to bound the evaluation error 
	
		$$\|f(\bfb)-\wtf(\wtbfb)\|$$
	in terms of
			$\delta_f \as \|f-\wtf\|$
		and
			$\delta_\bfb \as \|\bfb-\wtbfb\|$.
	This will be done by induction on $n$.
	Our analysis aims not just
	to produce some error bound, but to express this error
	in terms that are easily understood, and which reveals
	the underlying inductive structure.
	Towards this end, we introduce the following \dt{$\beta$-bound}
	function: if $d$ is a positive integer and $b\in \CC$, 
		
		\beql{beta}
		\beta(d,b)\as \sum_{i=0}^d |b|^i.
		\eeql
\moveToApp{
	A simple application of this $\beta$-bound is:
	\bleml{beta}
		Let $b\in\CC$ and $f\in\CC[z]$. 
		If $d$ is the degree of $f$, then
		
			$$|f(b)|\le \|f\|\cdot \beta(d, b),\qquad
			|f'(b)|\le d\|f\|\cdot \beta(d-1, b).$$
	\eleml
	Note that 
		$\beta(d,b)\le \max\set{d+1, \frac{|b|^{d+1}-1}{|b|-1}}$.
}
	We first treat the case $n=1$.  It will
	serve as the base for the inductive proof.
\moveToApp{%
	Its proof requires a complex version of the
	Mean Value Theorem.  Since this result is not well-known, we
	provide a statement and proof.

\bthmT{Complex Mean Value Theorem}{mvt}
	If $f:\CC\to\CC$ is holomorphic, then for
	any $a, b\in\CC$,
	
			$$f(b)-f(a) = \omega\cdot (b-a)\cdot f'(\xi)$$
	for some $\xi$ in the line segment $[a,b]$
	and some $\omega\in\CC$ with $|\omega|\le 1$.
\ethmT
}
\moveToApp{
\bpf
	This is a simple application of
	a similarly little known theorem of Darboux (1876)
	\cite{darboux:sur:87} which gives a
	finite Taylor expansion of $f$; 
	see B\"unger's formulation and proof in
	\cite[Appendix]{batra:effective:10}. 
	For any $k\ge 1$, the theorem says
	
		$$f(b)=\sum_{i=0}^{k-1} \frac{(b-a)^i}{i!}f^{(i)}(a)
			+ \omega \frac{(b-a)^k}{k!} f^{(k)}(\xi)$$
	for some $\xi$ in the line segment $[a,b]$ and $\omega\in\CC$
	with $|\omega|\le 1$.  Choosing $k=1$,
	$f(b)=f(a)+\omega(b-a)f'(\xi)$ or
	$f(b)-f(a)=\omega(b-a)f'(\xi)$.
	\epf
}

\moveToApp{
\bcorT{Complex Mean Value Inequality}{mvi}
	For all $a,b\in\CC$, there is some $\xi\in [a,b]$ such that
	
		$$|f(b)-f(a)| \le |b-a|\cdot |f'(\xi)|.$$
	\ecorT
}

\ignore{
	Let $\bfd$ be the degree sequence of $f$.
	For $i=1\dd n$, we want to bound
		$$\delta_k \as \|f(\bfb\subx[k]) -\wtf(\wtbfb\subx[k])\|.$$
	For instance, consider $\delta_1$.
	$f(\bfb\subx[1])\in \CC[z_2\dd z_n]$.
	
	THIS MAY CHANGE:
	Let us fix the following real parameters
		\beql{para}
		n,d,L,b
		\eeql
	assumed to be at least $1$.	
	A function of the form
		$$\beta=\beta_{n,d,L,b}:(\RR_{\ge 1})^2 \to \RR_{\ge 0}$$
	is called a \dt{evaluation bound} if	
	the following holds: suppose the polynomials
		$$f, \wtf \in \CC[\bfz] =\CC[z_1\dd z_n]$$
	and points
		$$\bfb, \wtbfb \in \CC^n$$
	satisfies 
	\bitem
	\item degree bound: $d\ge \max\set{\deg f, \deg \wtf}$
	\item height bound: $L \ge \|f\|$.
	    	Viewing $\wtf$ as a perturbation of $f$, 
	    		let $\delta_L \as \|f -\wtf\|$.
	\item domain bound: $b \ge \|\bfb\|$.
	    	Viewing $\wtbfb$ as a perturbation of $\bfb$,
	    		let $\delta_b \as \|\bfb -\wtbfb\|$.
	\eitem
	Here, we assume that $\|f\|$ is the infinity norm on polynomials
	(i.e., the maximum absolute value of the coefficients),
	and $\|b\|$ is the infinity norm on vectors
	(i.e., the maximum absolute value of its components).
	Also, let $\maxone (x) \as \max\set{1,|x|}$.
	We thus want to define $\beta$ to satisfy that
		\beql
		| \wtf(\wtbfb)- f(\bfb) | \le \beta (\| f-\wtf\|,
				\|\bfb -\wtbfb\|)
					=\beta(\delta_L, \delta_b).
		\eeql
}%

\moveToApp{
	\blemT{Case $n=1$}{n=1}\ \\
		Let $f,\wtf\in\CC[z]$, $b,\wtb\in\CC$, and
		$\bfd (\wtf)\le \bfd (f) \le d$.
		If $\wtf = f\pm\delta_f$ and $\wtb=b\pm \delta_b$, then:
	    \begin{center}
		\begin{tabular}{cllllll}
			(i) & $|f(b)-f(\wtb)|$
				& $\le$
				& $\delta_b\cdot \|f'\|\cdot \beta(d,|b|+\delta_b)$ 
				& where $f'$ is the differentiation of $f$.\\
			(ii) & $|f(\wtb)-\wtf(\wtb)|$
				& $\le$
				& $\delta_f \cdot \beta(d,|b|+\delta_b)$ & \\
			(iii) & $|f(b)-\wtf(\wtb)|$
				& $\le$
				& $\Big[ \delta_f + \delta_b \cdot \|f'\|\Big]\cdot
						\beta(d,|b|+\delta_b).$ &
		\end{tabular}
	    \end{center}
	\elemT
}
\moveToApp{
	\bpf
	\benum[(i)]
	\item
		By the complex mean value inequality (\refCor{mvi}):

		\beqarrays
		| f(b) -f(\wtb) |
			&\le& |b -\wtb| \cdot
					|f'(b\pm\delta_b)| \\
			&\le& \delta_b
				\cdot \sum_{i=1}^{d} \left| i f_i
					(|b|+\delta_b)^{i-1} \right| \\
			&\le& \delta_b
				\cdot \|f'\| \sum_{i=0}^{d-1}
					\left| (|b|+\delta_b)^i\right|.
		\eeqarrays
	\item Also
		\beqarrays
		| f(\wtb) -\wtf(\wtb) |
			&=& \Big|\sum_{i=0}^d (f_i-\wtf_i)\wtb^i\Big| \\
			&\le& \delta_f\cdot \sum_{i=0}^d \Big|\wtb^i\Big| \\
			&\le& \delta_f\cdot \beta(d,|b|+\delta_b).
		\eeqarrays
	\item This follows from the triangular inequality
		
		$$| f(b) - \wtf(\wtb) | \le 
			| f(b) -f(\wtb) |
			+ |f(\wtb) - \wtf(\wtb)|.$$
		and the bounds in parts (i) and (ii).
	\eenum
	\epf
}
	
	The appearance of $\|f'\|$ in the above bound
	may be replaced by $d\|f\|$.  Below, we develop similar
	bounds on partial derivatives in the multivariate case.
	\ignore{
	 Remark: our lemma could also be modified by using
	 an alternative triangular inequality:
	 
		$$| f(b) - \wtf(\wtb) | \le 
			| f(b) -\wtf(b) |
			+ |\wtf(b) - \wtf(\wtb)|.$$
	 I am not sure about the trade-offs but it seems
	 the version we choose is simpler.
	}%
	For a general $n>1$,
	we need to generalize the notations:
	
		\beql{fwtf}
		f,\wtf \in\CC[\bfz],\qquad \bfb,\wtbfb\in\CC^n
		\eeql
	satisfying 
	$\wtbfb=\bfb \pm \bfdelta\bfb$
		(i.e., $\wtbfb_i=\bfb_i \pm \bfdelta\bfb_i$ for each $i$).
	Let $\bfd = \bfd(f)$
		(i.e., $\bfd_i = \degree_{\bfz_i}(f)$ for each $i$).
	The \dt{support} of $f$ is $\supp(f)\ib\NN^n$ where
		$f = \sum_{\bfalpha\in \supp(f)} c_\bfalpha \bfz^\bfalpha$
	where $c_\bfalpha\in\CC\setminus \set{0}$. Here,
		$\bfz^\bfalpha\as \prod_{i=1}^n \bfz_i^{\bfalpha_i}$.
	We assume that $\supp(\wtf)\ib\supp(f)$.
	Our induction variable is $k=1\dd n$.
	For $\bfalpha\in\NN^n$, let
		$\pi_k(\bfalpha)\as (0\dd 0,\bfalpha_{k+1}\dd \bfalpha_n)$.
	E.g., if $k=n$ then $\pi_k(\bfalpha)=\0$.
	Thus $\bfalpha -\pi_k(\bfalpha) = 
		(\bfalpha_1\dd \bfalpha_k,0\dd 0)$.
	Next define
		$\supp_k(f)\as \set{\pi_k(\bfalpha): \bfalpha\in\supp(f)}$.
	With this notation, we can write
	
		\beql{fk}
		f = \sum_{\bfalpha\in \supp_k(f)}
			f_\bfalpha \bfz^\bfalpha
			\eeql
	where each $f_\bfalpha\in\CC[\bfz\subx[k]]$.
	E.g., if $k=n$ then $\supp_k(f)=\set{\0}$ and so $f_{\0}=f$.

	{\bf Running Example.} Consider
		
		\beql{f}
		f=xy+(x^3-1)y^2z + (x^2-y^2)z^3
		\eeql
	where $\bfz=(x,y,z)$.
	Then $\supp(f)=\set{110, 321, 021, 203, 023}$.
	We can represent $f$ using the support 
	$\supp_1(f)=\set{010, 021, 003, 023}$ as follows:
		$
		f = f_{010}\cdot y+f_{021}\cdot y^2z + f_{003}\cdot z^3
				f_{023}\cdot y^2z^3
		$
	where $f_{010}=x$, $f_{021}=x^3-1$, $f_{003}=x^2$, $f_{023}=-1$.
	Alternatively, using the support
	$\supp_2(f)=\set{000, 001, 003}$, we can write
		$
		f = f_{000}+f_{001}\cdot z + f_{003}\cdot z^3
		$
	where 
	$f_{000}=xy$, $f_{001}=(x^3-1)y^2$, $f_{003}=(x^2-y^2)$.

	Using \refeQ{fk}, the partial specialization
	$f(\bfb\subx[k])\in\CC[z_{k+1}\dd z_n]$
	may be written 
		
		$$f(\bfb\subx[k])
			= \sum_{\bfalpha\in \supp_k(f)}
			f_\bfalpha(\bfb\subx[k])\cdot \bfz^\bfalpha$$
	It follows that
		
		\beql{fb}
			\|f(\bfb\subx[k])\| 
				=\max_{\bfalpha\in \supp_k(f)}
				  \Big|f_\bfalpha(\bfb\subx[k])\Big|.
		\eeql
	The $k$-th partial derivative is
		$\del[k] f \as \frac{\partial f}{\partial z_k} 
		= \sum_{\bfalpha\in \supp_k(f)}
			(\del[k]f_\bfalpha) \bfz^\bfalpha.$
	Upon evaluation at $\bfb\subx[k]$,
	its norm is given by
		
		\beql{delfb}
			\|\del[k] f(\bfb\subx[k])\|
				=\max_{\bfalpha\in \supp_k(f)}
				\Big|\del[k] f_\bfalpha(\bfb\subx[k])\Big|.
		\eeql
	Using our running example \refeQ{fk}, let $k=2$.  Then
	$f=f_{000}+f_{001}\cdot z + f_{003}\cdot z^3$
	with $f_{001}=xy$, $f_{001}=(x^3-1)y^2$, $f_{003}=(x^2-y^2)$.
	Thus $\del[2] f = x + (x^3-1)2y\cdot z -2y\cdot z^3$.
	If $\bfb\subx[2]=(-1,3)$, then
	$\|f(\bfb\subx[2])\|=\max\set{3, 18, 8}=18$
	and
	$\|\del[2] f(\bfb\subx[2])\|=\max\set{1, 12, 6}=12$.

\moveToApp{%
	We are ready for the generalization of \refLem{n=1}.
}
	Assume that we are given $f,\wtf\in\CC[\bfz]=\CC[\bfz\subx]$
	and $\bfb,\wtbfb\in\CC$.  Also the degree sequences satisfies 
		$\bfd(\wtf)\le \bfd(f)$, that is the inequality holds
                componentwise.  Then we may define these quantities 
		for $k=1\dd n$:
		
		\beqarrys
			\delta_k \bfb &\as& 
				|\bfb_k -\wtbfb_k|, \\
			\delta_k f &\as& 
				\|f(\bfb\subx[k])-\wtf(\wtbfb\subx[k])\|
					& \qquad(\textrm{with } \delta_0 f=\|f-\wtf\|),\\
			\bfbeta_k &\as& 
				\beta(\bfd_k, \bfb_k) \\
			\wtbfbeta_k &\as& 
				\beta(\bfd_k, |\bfb_k| + \delta_k\bfb).
		\eeqarrys
		Note that $\delta_k$ is a operator that must
		attach to some function $f$ or vector $\bfb$ to
		denote the ``$k$th perturbation'' of $f$ or $\bfb$.
\moveToApp{%
	We may restate \refLem{n=1}(iii) using the new notations:
	\begin{coro}
		For a univariate $f$,
		
		\beql{n=1}
			\delta_1 f \le \Big[ \delta_0 f
			+ \delta_1 \bfb \cdot \bfd_1 \cdot\|f\|\Big]\wtbfbeta_1.
		\eeql
	\end{coro}
	We now address the case of multivariate $f$:
}

	\blemT{=Lemma 5 in Text}{n>=1}\ \\
		For $n\ge 1$ and each $k=1\dd n$:
			    \begin{center}
		\begin{tabular}{crlllll}
			(i) & $\| f(\bfb\subx[k])-f(\bfb\subx[k-1])(\wtbfb_k) \|$
				& $\le$
				& $\delta_k \bfb \cdot \|\del[k]f(\bfb\subx[k-1])\|
					\cdot \wtbfbeta_k$. \\
			(ii) & $\|f(\bfb\subx[k-1])(\wtbfb_k)-\wtf(\wtbfb\subx[k])\|$
				& $\le$
				& $\delta_{k-1} f \cdot \wtbfbeta_k$.\\
			(iii) & $\delta_k f$
				& $\le$
				& $\Big[ \delta_k \bfb \cdot \|\del[k]f(\bfb\subx[k-1])\|
					+ \delta_{k-1} f \Big]\cdot \wtbfbeta_k$.
		\end{tabular}
	    \end{center}
	\elemT
\moveToApp{
	\bpf
		We note that (iii) amounts to adding the inequalities
		of (i) and (ii): specifically,
			$\delta_k f 
			\le \| f(\bfb\subx[k])-f(\bfb\subx[k-1])(\wtbfb_k) \|
			+ \|f(\bfb\subx[k-1])(\wtbfb_k)-\wtf(\wtbfb\subx[k])\|$.
			Thus we only have to verify (i) and (ii).
		This will be shown by induction on $k$.

		Suppose $k=1$. 
		This will be an application of \refLem{n=1}(i) and (ii).
		We use the fact that
  		$f = \sum_{\bfalpha\in \supp_1(f)} f_\bfalpha \bfz^\bfalpha$,
		and $f(\bfb\subx[k-1])=f(\bfb\subx[0])=f$.
		Then (i) becomes
		
	\beqarrys
		\| f(\bfb_1)-f(\wtbfb_1) \|
			&=& \| \sum_{\bfalpha\in \supp_1(f)}
				(f_\bfalpha(\bfb_1)- f_\bfalpha(\wtbfb_1))
 					\bfz^\bfalpha \| \\
			&=& \max_{\bfalpha\in \supp_1(f)}
				|f_\bfalpha(\bfb_1)- f_\bfalpha(\wtbfb_1)| \\
			&\le& \max_{\bfalpha\in \supp_1(f)}
				\delta_1\bfb\cdot\|f_\bfalpha^'\| \cdot\wtbfbeta_1
					& (\textrm{by } \refLem{n=1}(i)) \\
			&=& \max_{\bfalpha\in \supp_1(f)}
				\delta_1\bfb\cdot\|\del[1] f_\bfalpha\|\cdot\wtbfbeta_1\\
			&=& \delta_1\bfb\cdot \|\del[1] f\|\cdot \wtbfbeta_1.
	\eeqarrys
	Similarly, (ii) follows from
	
 	\beqarrys
		\| f(\wtbfb_1)-\wtf(\wtbfb_1) \|
			&=& \|\sum_{\bfalpha\in \supp_1(f)}
				(f_\bfalpha(\wtbfb_1)- \wtf_\bfalpha(\wtbfb_1))
 				\bfz^\bfalpha \| \\
			&=& \max_{\bfalpha\in \supp_1(f)}
				\Big|f_\bfalpha(\wtbfb_1)-\wtf_\bfalpha(\wtbfb_1)\Big|\\
			&\le& \max_{\bfalpha\in \supp_1(f)}
				\delta_{f_\bfalpha}\cdot \wtbfbeta_1
				& (\textrm{by } \refLem{n=1}(ii)) \\
			&=& \|f -\wtf\|\cdot\wtbfbeta_1\\
			&=& \delta_0 f \cdot\wtbfbeta_1.
	\eeqarrys

		Suppose $k>1$.  We now prove (i).
		The left hand side (LHS)
		$\| f(\bfb\subx[k])-f(\bfb\subx[k-1])(\wtbfb_k) \|$
		is the maximum of
		\btableTwo[0.9]{
			$|f_\bfalpha(\bfb\subx[k])
				-f_\bfalpha(\bfb\subx[k-1])(\wtbfb_k)|$
			&&&& (A)}
		where $\bfalpha$ ranges over $\supp_k(f)$.
		We can rewrite (A) in the form
			$|f_\bfalpha(\bfb\subx[k-1])(\bfb_k)
				-f_\bfalpha(\bfb\subx[k-1])(\wtbfb_k)|$.
		Applying \refLem{n=1}(i), we can upper bound (A) by
			``$\delta_b\cdot \|f'\|\cdot \beta(d,|b|+\delta_b)$''
		where
			``$\delta_b$'' here is $|\bfb_k-\wtbfb_k|=\delta_k$,
			``$\|f'\|$'' is $\|\del[k]f_\bfalpha(\bfb\subx[k-1])\|$
				and 
			``$\beta(d,|b|+\delta_b)$'' is $\bfbeta_k$.
		This establishes (i).
		Finally (ii) is proved by a similar invocation of
		\refLem{n=1}(ii).
	\epf
}

	We now have a recursive bound $\|\delta_n f\|$.
	But we need to convert the bound to only depend on the
	data $\|\bfb\|, \|f\|, \delta_k\bfb$.
	In particular, we remove any occurrences of
	$\del[k]f_\bfalpha$ with the help of the next lemma:

	\blemT{= Lemma 6 in Text}{evalbound}
		For $k=1\dd n$:
		\ \\(i)
		$\|f(\bfb\subx[k])\|
			\le \|f\|\cdot \prod_{i=1}^k \bfbeta_i$
		\\(ii)
		For $\bfalpha\in\supp_k(f)$,
		\begin{center}
		\begin{tabular}{crlllll}
			$\Big\|\del[k]f_\bfalpha(\bfb\subx[k-1])\Big\|$
				& $\le$
				& $\bfd_k\cdot \|f_\bfalpha(\bfb\subx[k-1])\|$.
		\end{tabular}
		\end{center}
		(iii)
		$\|\del[k]f(\bfb\subx[k-1])\| \le \bfd_k
				\cdot \|f\|\cdot \prod_{i=1}^{k-1} \bfbeta_i$
	\elemT
\moveToApp{%
	\bpf
	(i) The LHS of the inequality is equal to the
		maximum of $|f_\bfalpha(\bfb\subx[k])|$
		where $\bfalpha\in\supp_k(f)$.
		First consider $k=1$.  In this case,
		$f_\bfalpha$ is a univariate polynomial in $\bfz_1$
		of degree at most $\bfd_1$,
		say $f_\bfalpha(\bfz_1)=\sum_{i=0}^{\bfd_1}c_i \bfz_1^i$
		where $c$ is a coefficient of $f$.  By \refLem{beta}, 
		
		$$|f_\bfalpha(\bfb_1)|
			\le \|f_\bfalpha\| \bfbeta_1
			\le \|f\| \bfbeta_1,$$
		proving the result for $k=1$.  For $k>1$,
		each $f_\bfalpha$ is a polynomial in $\bfz\subx[k]$,
		and we can write $f_\bfalpha(\bfb\subx[k])$
		as $f_\bfalpha(\bfb\subx[k-1])(\bfb_k)$.
		By induction, the polynomial $f_\bfalpha(\bfb\subx[k-1])(\bfz_k)$
		has norm at most 
		$\|f\|\cdot\prod_{i=1}^{k-1}\bfbeta_i$.
		Moreover, its degree is at most $\bfd_k$.  So evaluating
		it at $\bfb_k$ gives a value of size at most 
		$\|f\|\cdot\prod_{i=1}^{k}\bfbeta_i$.
		\\(ii)
		Write
		$f_\bfalpha(\bfb\subx[k-1])  = \sum_{i=0}^{\bfd_k} c_i \bfz_k^i$
		where $c_i\in\CC$ satisfies
		$|c_i|\le \|f_\bfalpha(\bfb\subx[k-1])\|$.
		\ignore{
		Then
			$$\Big|\del[k]f_\bfalpha(\bfb\subx[k])\Big|
				= \Big|\sum_{i=1}^{\bfd_k} i c_i \bfb_k^{i-1}\Big|
				\le \bfd_k \|f_\bfalpha(\bfb\subx[k-1])\|
			 		\cdot \beta(\bfd_k-1,\bfb_k).$$
		}%
		Thus $\del[k]f_\bfalpha(\bfb\subx[k-1])$  is a polynomial
		with norm
		
			$$\Big\|\del[k]f_\bfalpha(\bfb\subx[k-1])\Big\|
				\le \bfd_k \|f_\bfalpha(\bfb\subx[k-1])\|.$$
		\\(iii) Letting $\bfalpha$ range over $\supp_k(f)$,
		
		\beqarrys
			\|\del[k]f(\bfb\subx[k-1])\| &=& \max_{\bfalpha}
					\|\del[k]f_\bfalpha(\bfb\subx[k-1])\| &\\
				&\le& \max_{\bfalpha}
					\bfd_k \cdot \|f_\bfalpha(\bfb\subx[k-1])\|
                    & \qquad (\textrm{from part (ii) second formula})\\
				&\le& \bfd_k \cdot
						\max_{\bfalpha} \|f_\bfalpha(\bfb\subx[k-1])\|\\
				&\le& \bfd_k
						\cdot \|f(\bfb\subx[k-1])\| \\
				&\le& \bfd_k 
						\cdot \|f\|
						\cdot \prod_{i=1}^{k-1}\bfbeta_i
					& \qquad (\textrm{from part (i)})\\
		\eeqarrys
	\epf
}
	
	Putting it all together:
	\bthmT{=Theorem 2 in Text}{deltaf}
	For $k=1\dd n$,
	
	$$\delta_k f \le
			\Big[ \delta_0 f + \|f\|
				\cdot \sum_{i=1}^k \bfd_i\cdot \delta_i \bfb\Big]
				\cdot \Big(\prod_{i=1}^k \wtbfbeta_i\Big).$$
	\ethmT
\moveToApp{%
	\bpf
	When $k=1$, our formula is
	
	$$\delta_1 f \le \Big[ \delta_0 f + \|f\|
			\cdot \bfd_1\cdot \delta_1 \bfb \Big] \wtbfbeta_1. $$
	follows from the case $k=1$ of \refLem{n>=1}(iii).
	For $k>1$, we use induction:
		\begin{center}{\scriptsize
		\begin{tabular}{crlllll}
			$\delta_k f$
				& $\le$
				& $\Big[ \delta_k \bfb \cdot \|\del[k]f(\bfb\subx[k-1])\|
					+ \delta_{k-1} f \Big]\cdot \wtbfbeta_k$
				& \textrm{(By \refLem{n>=1}(iii)}\\
				& $\le$
				& $\Big[ \delta_k \bfb \cdot \|\del[k]f(\bfb\subx[k-1])\|
					+ \Big\{ \delta_0 f + \|f\|
					\cdot \sum_{i=1}^{k-1} \bfd_i\cdot \delta_i \bfb\Big\}
					\cdot \Big(\prod_{i=1}^{k-1} \wtbfbeta_i\Big)
					\Big] \cdot \wtbfbeta_k$
				& \textrm{(By induction)}\\
				& $\le$
				& $\Big[ \delta_k \bfb \cdot
					\bfd_k \cdot \|f\|
						\cdot\Big(\prod_{i=1}^{k-1} \bfbeta_i\Big)
					+ \Big\{ \delta_0 f + \|f\|
					\cdot \sum_{i=1}^{k-1} \bfd_i\cdot \delta_i \bfb\Big\}
					\cdot \Big(\prod_{i=1}^{k-1} \wtbfbeta_i\Big)
					\Big] \cdot \wtbfbeta_k$
				& \textrm{(By \refLem{evalbound}(iii))}\\
				& $\le$
				& $\Big[ \delta_k \bfb \cdot \bfd_k \cdot \|f\|
					+ \Big\{ \delta_0 f + \|f\|
					\cdot \sum_{i=1}^{k-1} \bfd_i\cdot \delta_i \bfb\Big\}
					\Big] \cdot \Big(\prod_{i=1}^{k} \wtbfbeta_i\Big)$
				& \textrm{(since $\bfbeta_i\le \wtbfbeta_i$)}\\
				& $=$
				& $\Big[ \delta_0 f + \|f\|
					\cdot \sum_{i=1}^{k} \bfd_i\cdot \delta_i \bfb \Big]
					\cdot \Big(\prod_{i=1}^{k} \wtbfbeta_i\Big)$.
		\end{tabular}}
		\end{center}
	\epf
}

	The next lemma answers the question:
	given $\delta_L>0$,
	how can we ensure that 
	
		$$\delta_{n-1}f \as
			\|f(\bfb\subx[n-1])-\wtf(\wtbfb\subx[n-1])\|$$
	is upper bounded by $\delta_L$?
	\blemT{=Lemma 7 in Text}{marc} \ \\
		\begin{tabular}{lrllll}
			\multicolumn{3}{l}{
				Given $\delta_L>0$,
				$f,\wtf\in\CC[\bfz]$ and
				$\bfb,\wtbfb\in\C^{\dim-1}$ where $\dim>1$.}\\
			\multicolumn{3}{l}{
				Let $d=\max(\degree_{\bfz_i}(f))$ and $M=\|\bfb\|+1$.}\\
	  	If\\
			& $\delta_f $
				& $\leq \frac{\delta_L}{2((d+1)M^d)^{n-1}}$
				& (*)\\
			and\\
			& $\delta_\bfb $
				& $\leq \min(1,
					\frac{\delta_L}{2d\|f\|(n-1)((d+1)M^d)^{n-1}})$,
				& (**)\\
			then\hspace*{3mm}\\
			& $\delta_{n-1} f$
				& $\leq \delta_L.$
		\end{tabular}
	\elemT
	
\moveToApp{%
	\bpf
		Note that $\delta_f \as \|f-\wtf\|$ in (*)
		and $\delta_\bfb \as \|\bfb-\wt\bfb\|$ in (**).
	Since $\delta_\bfb\leq 1$, we conclude that $\|\wt\bfb\|\leq M$.
	Using the bounds $\bfbeta_k\le \wtbfbeta_k\leq (d+1)M^d$,
		the bound of Theorem~\ref{thm:deltaf} for the case
		$k=n-1$ becomes

	\begin{eqnarray*}
	 	\delta_{n-1} f & \le &\Big(\prod_{i=1}^{n-1} \wtbfbeta_i\Big)
				\Big[ \delta_f + \|f\|
						\cdot \sum_{i=1}^{n-1} \Big( \bfd_i\cdot
	                               \delta_i \bfb \Big)
	                               \Big]\\
		 & \le & \Big((d+1)M^d \Big)^{n-1}
		         \Big[ \delta_f + d\|f\|\delta_\bfb (n-1)
		         \Big].
	\end{eqnarray*}
	The inequalities (*) on $\delta_f$, and (**) on $\delta_\bfb$,
	are designed to ensure that $\delta_{n-1} f \leq \delta_L$. 
	\epf
}

\ignore{
}%